\documentclass{article}%
\usepackage{amsmath}
\usepackage{amssymb}
\usepackage{amsfonts}
\usepackage{graphicx}%
\newtheorem{theorem}{Theorem}

\newtheorem{corollary}{Corollary}

\newtheorem{definition}{Definition}

\newtheorem{proposition}{Proposition}
\newtheorem{remark}{Remark}

\newenvironment{proof}[1][Proof]{\noindent\textbf{#1.} }{\ \rule{0.5em}{0.5em}}
\newcommand{\bpartial}{\mathop{\partial\kern -4pt\raisebox{.8pt}{$|$}}}
\newcommand{\bra}{\mathopen{[\kern-1.6pt[}}
\newcommand{\ket}{\mathclose{]\kern-1.5pt]}}
\newcommand{\bbra}{\mathopen{[\kern-2.2pt[\kern-2.3pt[}}
\newcommand{\bket}{\mathclose{]\kern-2.1pt]\kern-2.3pt]}}
\makeindex
\begin{document}

\title{The Square of the Dirac and spin-Dirac Operators on a Riemann-Cartan
Space(time)\thanks{to appear: \textit{Reports on Mathematical Physics}
\textbf{60}(1), 135-157 (2007).}}
\author{{\footnotesize E. A. Notte-Cuello} \linebreak\\{\footnotesize Departamento de Matem\'{a}ticas, Universidad de La Serena,}\\{\footnotesize Av. Cisterna 1200, La Serena-Chile.}\\{\small e-mail: enotte@userena.cl}
\and {\footnotesize W. A. Rodrigues Jr. }\\{\footnotesize Institute of Mathematics, Statistics and Scientific
Computation,}\\{\footnotesize IMECC-UNICAMP CP 6065, 13083-859 Campinas, SP, Brazil.}\\{\small e-mail: walrod@ime.unicamp.br}
\and {\footnotesize Q. A. G. Souza}\\$\hspace{-0.1cm}${\footnotesize Institute of Mathematics, Statistics and
Scientific Computation,}\\{\footnotesize IMECC-UNICAMP CP 6065, 13083-859 Campinas, SP, Brazil.}\\{\small quin@ime.unicamp.br}}
\maketitle

\begin{abstract}
In this paper we introduce the Dirac and spin-Dirac operators associated to a
connection on Riemann-Cartan space(time) and \ standard Dirac and spin-Dirac
operators associated with a Levi-Civita connection on a Riemannian
(Lorentzian) space(time) and calculate the square of these operators, which
play an important role in several topics of modern Mathematics, in particular
in the study of the geometry of moduli spaces of a class of black holes, the
geometry of NS-5 brane solutions of type II supergravity theories and BPS
solitons in some string theories. We obtain a generalized Lichnerowicz
formula, decompositions of the Dirac and spin-Dirac operators and their
squares in terms of the \textit{standard }Dirac and spin-Dirac operators and
using the fact that spinor fields (sections of a spin-Clifford bundle) have

representatives in the Clifford bundle we present also a noticeable relation
involving the spin-Dirac and the Dirac operators.

\end{abstract}


\section{Introduction}

Recently, in several applications of theoretical physics and differential
geometry, in a way or another the Dirac operator and its square on a
Riemann-Cartan space(time) has been used. In, e.g., \cite{Ra1} Rapoport
proposed to give a Clifford bundle approach to his theory of generalized
Brownian motion; in \cite{AT1} Agricola and Friedrich investigate the holonomy
group of a linear metric connection with skew-symmetric torsion and in
\cite{AT2} they introduced also an elliptic, second-order operator acting on a
spinor field, and in the case of a naturally reductive space they calculated
the Casimir operator of the isometry group. The square of the spin-Dirac
operator also appears naturally in the study of the geometry of moduli spaces
of a class of black holes, the geometry of NS-5 brane solutions of type II
supergravity theories and BPS solitons in some string theories
\emph{(\cite{dalakov})} and many other important topics of modern mathematics
(see \cite{bismut,friedrich}). Some of the works just quoted present extremely
sophisticated and really complicated calculations and sometimes even erroneous ones.

This brings to mind that a simple theory of Dirac operators and their squares
acting on sections of the Clifford and Spin-Clifford bundles on Riemann-Cartan
space(times) has been presented in \cite{SR}, and further developed
in~\cite{rodoliv2006}. Using that theory, in Section 2 we first introduce the
standard Dirac operator $\mathop{\partial\kern -4pt\raisebox{.8pt}{$|$}}$
(associated with a Levi-Civita connection $D$ of a metric field $\mathbf{g}$)
acting on sections of the Clifford bundle of differential forms $\mathcal{C}%
\ell\left(  M,\mathtt{g}\right)  $ and next, in section 2.1 we introduce the
Dirac operator $\mbox{\boldmath$\partial$}$ (associated with an arbitrary
connection $\nabla$) and also acting on sections of $\mathcal{C}\ell\left(
M,\mathtt{g}\right)  $. Next, we calculate in Section 2.2 in a simple and
direct way the square of the Dirac operator on Riemann-Cartan space, and then
specialize the result for the simplest case of a scalar function $f\in\sec%
{\displaystyle\bigwedge\nolimits^{0}}
T^{\ast}M\hookrightarrow\sec$\texttt{ }$\mathcal{C}\ell\left(  M,\mathtt{g}%
\right)  $ in order to compare our results with the ones presented in
\cite{Ra1}. We give two calculations, one using the decomposition of the Dirac
operator into the \emph{standard} Dirac operator plus a term depending on the
torsion tensor (see Eq.(\ref{f17}) below) and another one, which follows
directly from the definition of the Dirac operator without using the standard
Dirac operator. Next we present a relation between the square of the Dirac and
the standard Dirac operators (acting on a \emph{scalar} function) in terms of
the torsion tensor and investigate also in Section 2.3 the relation between
those operators in the case of a null strain tensor. In Section 3, we present
a brief summary of the theory of the Spin-Clifford bundles ($\mathcal{C}%
\ell_{\mathrm{Spin}_{1,3}^{e}}^{\ell}(M,\mathtt{g})$ and $\mathcal{C}%
\ell_{\mathrm{Spin}_{1,3}^{e}}^{r}(M,\mathtt{g})$) and their sections (spinor
fields) and their representatives in a Clifford bundle ($\mathcal{C}%
\ell(M,\mathtt{g})$) following \cite{moro,rodoliv2006}. We recall in Section
3.1 some important formulas from the general theory of the covariant
derivatives of Clifford and spin-Clifford fields and in Section 3.2 we recall
the definition of the spin-Dirac operator $\mbox{\boldmath$\partial$}^{s}$
(associated with a Riemann-Cartan connection $\nabla$) acting on sections of a
spin-Clifford bundle . In section 3.3 we introduce the \emph{representatives}
of spinor fields in the Clifford \emph{bundle} and the important concept of
the representative of $\mbox{\boldmath$\partial$}^{s}$ (denoted
$\mbox{\boldmath$\partial$}^{(s)}$) that acts on the representatives of spinor
fields (see \cite{moro,rodoliv2006} for details). To make clear the
similarities and differences between $\mathcal{C}\ell(M,\mathtt{g})$ and
$\mathcal{C}\ell_{\mathrm{Spin}_{1,3}^{e}}^{\ell}(M,\mathtt{g})$, we write, in
Section 4, Maxwell \emph{equation} in both formalisms. In Section 5.1 we first
find the commutator of the covariant derivative of spinor fields on a
Riemann-Cartan space(time) and compare our result with one that can be found
in \cite{ramond}, which seems to neglect a term. Next in Section 5.2 we
calculate the square of the spin-Dirac operator on a Riemann-Cartan spacetime
and find a generalized Lichnerowicz formula. In Section 6, taking advantage
that any $\psi\in\sec\mathcal{C}\ell_{\mathrm{Spin}_{1,3}^{e}}^{\ell
}(M,\mathtt{g})$ can be written as $\psi=A1_{\Xi}^{\ell}$ with $A\in
\sec\mathcal{C}\ell(M,\mathtt{g})$ and $1_{\Xi}^{\ell}\in\sec\mathcal{C}%
\ell_{\mathrm{Spin}_{1,3}^{e}}^{\ell}(M,\mathtt{g})$ we find two noticeable
formulas: the first relates the square of the \emph{spin-Dirac operator}
($\theta^{\mathbf{a}}\nabla_{\mathbf{e}_{\mathbf{a}}}^{s}$) acting on $\psi$
with the square of the Dirac operator ($\theta^{\mathbf{a}}\nabla
_{\mathbf{e}_{\mathbf{a}}}$) acting on $A$; the second formula relates the
square of the \emph{spin-Dirac operator} ($\theta^{\mathbf{a}}\nabla
_{\mathbf{e}_{\mathbf{a}}}^{s}$) acting on $\psi$ with the square of the
standard Dirac operator ($\theta^{\mathbf{a}}D_{\mathbf{e}_{\mathbf{a}}}$). In
Section 7 we present our conclusions.

\section{The Standard Dirac Operator}

Let $M$ be a smooth differentiable manifold, $\mathbf{g}\in\sec T_{2}^{0}M$ a
smooth metric field, $\nabla$ a connection and $\mathbf{T}$ and $\mathbf{R}$,
respectively the torsion and curvature tensors of the connection $\nabla.$

\begin{definition}
Given a triple $\left(  M,\mathbf{g},\nabla\right)  :$

\begin{description}
\item[a)] it is called a Riemann-Cartan space if and only if%
\[
\nabla\mathbf{g}=0\qquad\text{and\qquad}\mathbf{T}\left[  \nabla\right]
\neq0.
\]

\item[b)] it is called a Riemann space if and only if%
\[
\nabla\mathbf{g}=0\qquad\text{and\qquad}\mathbf{T}\left[  \nabla\right]  =0.
\]
For each metric tensor defined on the manifold $M$ there exists one and only
one connection that satisfies these conditions. It is called the
\textit{Levi-Civita connection }of the metric considered and is denoted by
$D.$\ When $\dim M=4$ and the metric \textbf{$\mathtt{g}$} has signature
$\left(  1,3\right)  $ the triple $\left(  M,\mathbf{g},\nabla\right)  $ is
called a Riemann-Cartan spacetime and the triple $\left(  M,\mathbf{g}%
,D\right)  $ a Lorentzian spacetime\footnote{We recall (see, e.g.,
\cite{rodoliv2006}) that a Riemann-Cartan or a Lorentzian spacetime must be
orientable and time orientable.}.

\item[c)] it is called a Riemann-Cartan-Weyl space if and only if%
\[
\nabla\mathbf{g}\neq0\qquad\text{and\qquad}\mathbf{T}\left[  \nabla\right]
\neq0.
\]

\end{description}
\end{definition}

For the computation of the square of the Dirac operator on a Riemann-Cartan
space, we need first to introduce on the Clifford bundle of differential form
$\mathcal{C}\ell\left(  M,\mathtt{g}\right)  $ a differential operator
$\mathop{\partial\kern -4pt\raisebox{.8pt}{$|$}}$, called the \textit{standard
Dirac operator} \cite{rodoliv2006},\ which is associated with the Levi-Civita
connection of the Riemannian (or Lorentzian) structure $\left(  M,\mathbf{g}%
,D\right)  $. A Lorentzian spacetime for which $\mathbf{R}=0$ is called a
Minkowski spacetime. Note that we denoted by $\mathtt{g}\in\sec T_{0}^{2}M$
the metric tensor of the cotangent bundle.

Given $\mathbf{u}\in\sec TM$ and $u\in\sec%
{\displaystyle\bigwedge\nolimits^{1}}
T^{\ast}M\hookrightarrow\sec$\texttt{ }$\mathcal{C}\ell\left(  M,\mathtt{g}%
\right)  $ consider the tensorial mapping $A\longmapsto uD_{\mathbf{u}}A,$
$A\in\sec\mathcal{C}\ell\left(  M,\mathtt{g}\right)  .$ Since $D_{\mathbf{u}%
}J_{\mathtt{g}}\subseteq J_{\mathtt{g}},$ where $J_{\mathtt{g}}$ is the ideal
used in the definition of $\mathcal{C}\ell\left(  M,\mathtt{g}\right)  $ (see,
e.g., \cite{rodoliv2006} for details), the notion of covariant derivative
(related to the Levi-Civita connection) pass to the quotient bundle\texttt{
}$\mathcal{C}\ell\left(  M,\mathtt{g}\right)  .$

Let $U\subset M$ an open set and $\left\{  \mathbf{e}_{\alpha}\right\}  $ on
$TU\subset TM$ a moving frame with dual moving frame $\left\{  \theta^{\alpha
}\right\}  $, where $\theta^{\alpha}\in\sec%
{\displaystyle\bigwedge\nolimits^{1}}
T^{\ast}M\hookrightarrow\sec$\texttt{ }$\mathcal{C}\ell\left(  M,\mathtt{g}%
\right)  $.

\begin{definition}
The standard Dirac operator is the first order differential operator
\begin{equation}
\mathop{\partial\kern -4pt\raisebox{.8pt}{$|$}}=\theta^{\alpha}D_{\mathbf{e}%
_{\alpha}}. \label{f1}%
\end{equation}

\end{definition}

For $A\in\sec$\texttt{ }$\mathcal{C}\ell\left(  M,\mathtt{g}\right)  ,$%
\[
\mathop{\partial\kern
-4pt\raisebox{.8pt}{$|$}}A=\theta^{\alpha}\left(  D_{\mathbf{e}_{\alpha}%
}A\right)  =\theta^{\alpha}\mathbin\lrcorner\left(  D_{\mathbf{e}_{\alpha}%
}A\right)  +\theta^{\alpha}\wedge\left(  D_{\mathbf{e}_{\alpha}}A\right)
\]
and then we define%
\begin{equation}%
\begin{array}
[c]{ll}%
\mathop{\partial\kern -4pt\raisebox{.8pt}{$|$}}\mathbin\lrcorner A &
=\theta^{\alpha}\mathbin\lrcorner\left(  D_{\mathbf{e}_{\alpha}}A\right) \\
\mathop{\partial\kern -4pt\raisebox{.8pt}{$|$}}\wedge A & =\theta^{\alpha
}\wedge\left(  D_{\mathbf{e}_{\alpha}}A\right)
\end{array}
\label{f2}%
\end{equation}
in order to have%
\[
\mathop{\partial\kern
-4pt\raisebox{.8pt}{$|$}}=\mathop{\partial\kern
-4pt\raisebox{.8pt}{$|$}}\mathbin\lrcorner+\mathop{\partial\kern
-4pt\raisebox{.8pt}{$|$}}\wedge.
\]

\begin{proposition}
The standard Dirac operator $\mathop{\partial\kern
-4pt\raisebox{.8pt}{$|$}}$ is related to the exterior derivative $d$ and to
the Hodge codifferential $\delta$ by%
\[
\mathop{\partial\kern -4pt\raisebox{.8pt}{$|$}}=d-\delta,
\]
that is, we have $\mathop{\partial\kern
-4pt\raisebox{.8pt}{$|$}}\wedge=d$ and $\mathop{\partial\kern
-4pt\raisebox{.8pt}{$|$}}\mathbin\lrcorner=-\delta.$ For proof see, e.g.,
\cite{rodoliv2006}.
\end{proposition}

\subsection{The Dirac Operator in Riemann-Cartan Space}

We now consider a Riemann-Cartan-Weyl structure $\left(  M,\mathtt{g}%
,\nabla\right)  $ where $\nabla$ is an arbitrary linear connection, which in
general, is not metric compatible. In this genral case, the notion of
covariant derivative does \textit{not} pass to the quotient bundle
$\mathcal{C}\ell\left(  M,\mathtt{g}\right)  $ \cite{Cr}. Despite this fact,
it is still a \textit{well defined} operation and in analogy with the earlier
section, we can associate with it, acting on the sections of the Clifford
bundle $\mathcal{C}\ell\left(  M,\mathtt{g}\right)  $, the operator
\begin{equation}
\mbox{\boldmath$\partial$}=\theta^{\alpha}\nabla_{\mathbf{e}_{\alpha}},
\label{f3}%
\end{equation}
where $\left\{  \theta^{\alpha}\right\}  $ is a moving frame on $T^{\ast}U,$
dual to the moving frame $\left\{  \mathbf{e}_{\alpha}\right\}  $ on
$TU\subset TM$.

\begin{definition}
The operator $\mbox{\boldmath$\partial$}$ is called the Dirac operator (or
Dirac derivative, or sometimes the gradient) acting on sections of the
Clifford bundle.
\end{definition}

We also define%
\begin{equation}%
\begin{array}
[c]{ll}%
\mbox{\boldmath$\partial$}\mathbin\lrcorner A & =\theta^{\alpha}%
\mathbin\lrcorner\left(  \nabla_{\mathbf{e}_{\alpha}}A\right)  ,\\
\mbox{\boldmath$\partial$}\wedge A & =\theta^{\alpha}\wedge\left(
\nabla_{\mathbf{e}_{\alpha}}A\right)  ,
\end{array}
\label{f4}%
\end{equation}%
\begin{equation}
\mbox{\boldmath$\partial$}={\mbox{\boldmath$\partial$}\mathbin\lrcorner
}+{\mbox{\boldmath$\partial$}\wedge}. \label{f5}%
\end{equation}
The operator $\mbox{\boldmath$\partial$}\wedge$ satisfies \cite{rodoliv2006},
for every $A,B\in\sec$\texttt{ }$\mathcal{C}\ell\left(  M,\mathtt{g}\right)  $%
\[
\mbox{\boldmath$\partial$}\wedge\left(  A\wedge B\right)  =\left(
\mbox{\boldmath$\partial$}\wedge A\right)  \wedge B+\widehat{A}\wedge\left(
\mbox{\boldmath$\partial$}\wedge B\right)  ,
\]
where $\ \widehat{A}$ \ denote the main involution (or graded involution) of
$A\in\sec$\texttt{ }$\mathcal{C}\ell\left(  M,\mathtt{g}\right)  $.

Properties of this general operator are studied in \cite{rodoliv2006}.
Hereafter we suppose that\textbf{ }$\nabla$ is metric compatible, i.e.,
$\left(  M,\mathbf{\mathtt{g}},\nabla\right)  $ is Riemann-Cartan space(time),
and of course in this case $\nabla$ defines a connection in $\mathcal{C}%
\ell\left(  M,\mathtt{g}\right)  .$

Let $D_{\mathbf{e}_{\beta}}\theta^{\alpha}=-\mathring{\Gamma}_{\beta\rho
}^{\alpha}\theta^{\rho},$ and $\nabla_{\mathbf{e}_{\beta}}\theta^{\alpha
}=-\Gamma_{\beta\rho}^{\alpha}\theta^{\rho},$ where the covariant derivative
$\nabla_{\mathbf{e}_{\beta}}$ (which is now a $\mathbf{g}$-compatible
connection), has a non-zero torsion tensor\ whose components in the basis
\textbf{\ }$\{\mathbf{e}_{\alpha}\otimes\theta^{\beta}\otimes\theta^{\rho}\}$
are $T_{\beta\rho}^{\alpha}\equiv\Gamma_{\beta\rho}^{\alpha}-\Gamma_{\rho
\beta}^{\alpha}-c_{\beta\rho}^{\alpha}$.

\begin{proposition}
Let $\Theta^{\rho}=\frac{1}{2}T_{\alpha\beta}^{\rho}\theta^{\alpha}%
\wedge\theta^{\beta}\in\sec%
{\displaystyle\bigwedge\nolimits^{2}}
T^{\ast}M\hookrightarrow\sec$\texttt{ }$\mathcal{C}\ell\left(  M,\mathtt{g}%
\right)  $ the torsion $2$-forms of the connection $\nabla$ in an arbitrary
moving frame $\left\{  \theta^{\alpha}\right\}  .$ Then%
\begin{equation}%
\begin{array}
[c]{cc}%
\mbox{\boldmath$\partial$}\mathbin\lrcorner &
=\mathop{\partial\kern -4pt\raisebox{.8pt}{$|$}}\mathbin\lrcorner-\Theta
^{\rho}\mathbin\lrcorner\mathbf{j}_{\rho}\\
\mbox{\boldmath$\partial$}\wedge &
=\mathop{\partial\kern -4pt\raisebox{.8pt}{$|$}}\wedge-\Theta^{\rho
}\mathbin\lrcorner\mathbf{i}_{\rho},
\end{array}
\label{f6}%
\end{equation}
where $\mathbf{i}_{\rho}A=\theta_{\rho}\mathbin\lrcorner A,$ $\mathbf{j}%
_{\rho}A=\theta_{\rho}\wedge A,$ for every $A\in\sec$\texttt{ }$\mathcal{C}%
\ell\left(  M,\mathtt{g}\right)  $.\ For the proof, see
\emph{\cite{rodoliv2006}}.
\end{proposition}

\begin{proposition}
Let $\Theta^{\rho}=\frac{1}{2}T_{\alpha\beta}^{\rho}\theta^{\alpha}%
\wedge\theta^{\beta}\in\sec%
{\displaystyle\bigwedge\nolimits^{2}}
T^{\ast}M\hookrightarrow\sec$\texttt{ }$\mathcal{C}\ell\left(  M,\mathtt{g}%
\right)  $ the torsion $2$-forms and $f\in\sec%
{\displaystyle\bigwedge\nolimits^{0}}
T^{\ast}M\hookrightarrow\sec$\texttt{ }$\mathcal{C}\ell\left(  M,\mathtt{g}%
\right)  $,\ a scalar function, then%
\begin{equation}
\Theta^{\rho}\mathbin\lrcorner\left(  \theta_{\rho}\wedge\left(
\mathop{\partial\kern -4pt\raisebox{.8pt}{$|$}}f\right)  \right)
=-T_{\alpha\beta}^{\alpha}\mathbf{e}^{\beta}\left(  f\right)  . \label{f7}%
\end{equation}

\end{proposition}

\begin{proof}
From the Eq.(\ref{f1}) we have%
\begin{equation}
\Theta^{\rho}\mathbin\lrcorner\left(  \theta_{\rho}\wedge\left(
\mathop{\partial\kern -4pt\raisebox{.8pt}{$|$}}f\right)  \right)  =\frac{1}%
{2}T_{\alpha\beta}^{\rho}\left(  \theta^{\alpha}\wedge\theta^{\beta}\right)
\mathbin\lrcorner\left(  \theta_{\rho}\wedge\theta^{\delta}D_{\mathbf{e}%
_{\delta}}\left(  f\right)  \right)  \label{f8}%
\end{equation}
and recalling that for any $X_{k},Y_{k}\in\sec%
{\displaystyle\bigwedge\nolimits^{k}}
T^{\ast}M\hookrightarrow\sec$\texttt{ }$\mathcal{C}\ell\left(  M,\mathtt{g}%
\right)  $, $X_{k}\mathbin\lrcorner Y_{k}=\widetilde{X}_{k}\cdot Y_{k}%
=Y_{k}\llcorner X_{k}=X_{k}\cdot\widetilde{Y}_{k}$ (see, e.g.,
\cite{rodoliv2006}), where $\widetilde{X}_{k}$ denote the \textit{reversion}
operator of $X_{k}$, \ we can write,%

\[
\left(  \theta^{\alpha}\wedge\theta^{\beta}\right)  \mathbin\lrcorner\left(
\theta_{\rho}\wedge\theta^{\delta}\right)  =-\left(  \delta_{\rho}^{\alpha
}g^{\beta\delta}-\delta_{\rho}^{\beta}g^{\alpha\delta}\right)  .
\]
Then, Eq.(\ref{f8}) we get after some algebra%
\[
\Theta^{\rho}\mathbin\lrcorner\left(  \theta_{\rho}\wedge\left(
\mathop{\partial\kern -4pt\raisebox{.8pt}{$|$}}f\right)  \right)
=-T_{\rho\beta}^{\rho}\mathbf{e}^{\beta}\left(  f\right)
\]
and Eq.(\ref{f7}) is proved.
\end{proof}

\begin{proposition}
Let $f\in\sec%
{\displaystyle\bigwedge\nolimits^{0}}
T^{\ast}M\hookrightarrow\sec$\texttt{ }$\mathcal{C}\ell\left(  M,\mathtt{g}%
\right)  $ a scalar function, $d$ and $\delta$, respectively the exterior
derivative and the Hodge codifferential, then%
\begin{equation}
-\delta df=g^{\beta\alpha}\nabla_{\mathbf{e}_{\beta}}\nabla_{\mathbf{e}%
_{\alpha}}f-g^{\beta\rho}\mathring{\Gamma}_{\beta\rho}^{\alpha}\mathbf{e}%
_{\alpha}\left(  f\right)  \label{f9}%
\end{equation}

\end{proposition}

\begin{proof}
Using the Eqs. (\ref{f2}) we have%
\begin{align*}
-\delta df &
=\mathop{\partial\kern -4pt\raisebox{.8pt}{$|$}}\mathbin\lrcorner\left(
\mathop{\partial\kern -4pt\raisebox{.8pt}{$|$}}\wedge f\right)
=\mathop{\partial\kern -4pt\raisebox{.8pt}{$|$}}\mathbin\lrcorner\left(
\theta^{\alpha}\wedge D_{\mathbf{e}_{\alpha}}f\right)
=\mathop{\partial\kern -4pt\raisebox{.8pt}{$|$}}\mathbin\lrcorner\left(
\theta^{\alpha}\mathbf{e}_{\alpha}\left(  f\right)  \right)  \medskip\\
&  =\theta^{\beta}\mathbin\lrcorner\left(  D_{\mathbf{e}_{\beta}}%
\theta^{\alpha}\mathbf{e}_{\alpha}\left(  f\right)  \right)  =\theta^{\beta
}\mathbin\lrcorner\left(  \mathbf{e}_{\beta}\left(  \mathbf{e}_{\alpha
}f\right)  \theta^{\alpha}+\left(  D_{\mathbf{e}_{\beta}}\theta^{\alpha
}\right)  \mathbf{e}_{\alpha}\left(  f\right)  \right)  \medskip\\
&  =g^{\beta\alpha}\nabla_{\mathbf{e}_{\beta}}\nabla_{\mathbf{e}_{\alpha}%
}f-g^{\beta\rho}\mathring{\Gamma}_{\beta\rho}^{\alpha}\mathbf{e}_{\alpha
}\left(  f\right)
\end{align*}
and Eq.(\ref{f9}) is proved.
\end{proof}

\subsection{The Square of the Dirac Operator on a Riemann-Cartan Space(time)}

Let us now compute the square of the Dirac operator on a Riemann-Cartan
space(time). We have by definition,
\[%
\begin{array}
[c]{ll}%
\mbox{\boldmath$\partial$}^{2} & =\left(
{\mbox{\boldmath$\partial$}\mathbin\lrcorner}%
+{\mbox{\boldmath$\partial$}\wedge}\right)  \left(
{\mbox{\boldmath$\partial$}\mathbin\lrcorner}%
+{\mbox{\boldmath$\partial$}\wedge}\right) \\
& ={\mbox{\boldmath$\partial$}\mathbin\lrcorner}%
{\mbox{\boldmath$\partial$}\mathbin\lrcorner}%
+{\mbox{\boldmath$\partial$}\mathbin\lrcorner}%
{\mbox{\boldmath$\partial$}\wedge}+{\mbox{\boldmath$\partial$}\wedge
}{\mbox{\boldmath$\partial$}\mathbin\lrcorner}%
+{\mbox{\boldmath$\partial$}\wedge}{\mbox{\boldmath$\partial$}\wedge}%
\end{array}
\]
and writing
\[
\mathcal{L}_{+}={\mbox{\boldmath$\partial$}\mathbin\lrcorner}%
{\mbox{\boldmath$\partial$}\wedge}+{\mbox{\boldmath$\partial$}\wedge
}{\mbox{\boldmath$\partial$}\mathbin\lrcorner},
\]
we get
\begin{equation}
\mbox{\boldmath$\partial$}^{2}={\mbox{\boldmath$\partial$}^{2}%
\mathbin\lrcorner}+\mathcal{L}_{+}+{\mbox{\boldmath$\partial$}^{2}\wedge}.
\label{f10}%
\end{equation}
The operator $\mathcal{L}_{+}$ when applied to a scalar function corresponds,
for the case of a Riemann-Cartan space, to the wave operator introduced by
Rapoport \cite{Ra2} in his theory of Stochastic Mechanics. Obviously, for the
case of the standard Dirac operator, $\mathcal{L}_{+}$ reduces to the usual
Hodge Laplacian of the manifold \cite{SR,rodoliv2006}.

Let us now compute the square of the Dirac operator on a scalar function
$f\in\sec%
{\displaystyle\bigwedge\nolimits^{0}}
T^{\ast}M\hookrightarrow\sec$\texttt{ }$\mathcal{C}\ell\left(  M,\mathtt{g}%
\right)  $ using Eq.(\ref{f10}).

First we calculate $\mathcal{L}_{+}f$, which needs the calculation of $\left(
\mbox{\boldmath$\partial$}\mathbin\lrcorner\mbox{\boldmath$\partial$}\wedge
\right)  f$ and $\left(  \mbox{\boldmath$\partial$}\wedge
\mbox{\boldmath$\partial$}\mathbin\lrcorner\right)  f$.

\begin{description}
\item[a)] Using Eqs.(\ref{f6}) and Proposition 3, we have%
\begin{align}
\left(  \mbox{\boldmath$\partial$}\mathbin\lrcorner
\mbox{\boldmath$\partial$}\wedge\right)  f &
=\mbox{\boldmath$\partial$}\mathbin\lrcorner\left(
\mathop{\partial\kern -4pt\raisebox{.8pt}{$|$}}\wedge f-\Theta^{\rho}\wedge
i_{\rho}f\right)  =\mbox{\boldmath$\partial$}\mathbin\lrcorner\left(
\mathop{\partial\kern -4pt\raisebox{.8pt}{$|$}}\wedge f-\Theta^{\rho}%
\wedge\left(  \theta_{\rho}\mathbin\lrcorner f\right)  \right)  \nonumber\\
&  =\mbox{\boldmath$\partial$}\mathbin\lrcorner\left(
\mathop{\partial\kern -4pt\raisebox{.8pt}{$|$}}\wedge f\right)  =\left(
\mathop{\partial\kern -4pt\raisebox{.8pt}{$|$}}\mathbin\lrcorner-\Theta^{\rho
}\mathbin\lrcorner j_{\rho}\right)  \left(
\mathop{\partial\kern -4pt\raisebox{.8pt}{$|$}}\wedge f\right)  \nonumber\\
&  =\mathop{\partial\kern -4pt\raisebox{.8pt}{$|$}}\mathbin\lrcorner\left(
\mathop{\partial\kern -4pt\raisebox{.8pt}{$|$}}\wedge f\right)  -\Theta^{\rho
}\mathbin\lrcorner\left(  \theta_{\rho}\wedge\left(
\mathop{\partial\kern -4pt\raisebox{.8pt}{$|$}}\wedge f\right)  \right)
\nonumber\\
&  =-\delta df-\Theta^{\rho}\mathbin\lrcorner\left(  \theta_{\rho}%
\wedge\left(  \mathop{\partial\kern -4pt\raisebox{.8pt}{$|$}}f\right)
\right)  .\label{f11}%
\end{align}
Then, substituting the Eqs.(\ref{f9}) and (\ref{f7}) into Eq.(\ref{f11}) we
obtain
\begin{equation}
\left(  \mbox{\boldmath$\partial$}\mathbin\lrcorner
\mbox{\boldmath$\partial$}\wedge\right)  f=g^{\beta\alpha}\nabla
_{\mathbf{e}_{\beta}}\nabla_{\mathbf{e}_{\alpha}}f-g^{\beta\rho}%
\mathring{\Gamma}_{\beta\rho}^{\alpha}\mathbf{e}_{\alpha}\left(  f\right)
+T_{\alpha\beta}^{\alpha}\mathbf{e}^{\beta}\left(  f\right)  \label{f12}%
\end{equation}

\item[b)] Now, using Eq.(\ref{f6}) we have%
\begin{equation}
\left(  \mbox{\boldmath$\partial$}\wedge
\mbox{\boldmath$\partial$}\mathbin\lrcorner\right)
f=\mbox{\boldmath$\partial$}\wedge\left(
\mathop{\partial\kern -4pt\raisebox{.8pt}{$|$}}\mathbin\lrcorner
f-\Theta^{\rho}\mathbin\lrcorner j_{\rho}f\right)
=\mbox{\boldmath$\partial$}\wedge\left(  \theta^{\alpha}\mathbin\lrcorner
\mathbf{e}_{\alpha}\left(  f\right)  -\Theta^{\rho}\mathbin\lrcorner\left(
\theta_{\rho}\wedge f\right)  \right)  =0.\label{f13}%
\end{equation}
So, from the Eqs (\ref{f12}) and (\ref{f13}), we obtain%
\begin{equation}
\mathcal{L}_{+}f=g^{\beta\alpha}\nabla_{\mathbf{e}_{\beta}}\nabla
_{\mathbf{e}_{\alpha}}f-g^{\beta\rho}\mathring{\Gamma}_{\beta\rho}^{\alpha
}\mathbf{e}_{\alpha}\left(  f\right)  +T_{\alpha\beta}^{\alpha}\mathbf{e}%
^{\beta}\left(  f\right)  .\label{f14}%
\end{equation}

\item[c)] On the other hand, the first term of Eq.(\ref{f10}) is zero, i.e.,
$\mbox{\boldmath$\partial$}\mathbin\lrcorner
\mbox{\boldmath$\partial$}\mathbin\lrcorner f=0,$ because
$\mbox{\boldmath$\partial$}\mathbin\lrcorner f=\theta^{\alpha}%
\mathbin\lrcorner\nabla_{\mathbf{e}_{\alpha}}f=\theta^{\alpha}%
\mathbin\lrcorner\mathbf{e}_{\alpha}f=0$.

\item[d)] Now, using again Eq.(\ref{f6}) we calculate the last term of the
Eq.(\ref{f10}),%
\begin{align}
\mbox{\boldmath$\partial$}\wedge\mbox{\boldmath$\partial$}\wedge f &
=\mbox{\boldmath$\partial$}\wedge\left(
\mathop{\partial\kern -4pt\raisebox{.8pt}{$|$}}\wedge f-\Theta^{\rho}\wedge
i_{\rho}f\right)  =\mbox{\boldmath$\partial$}\wedge\left(
\mathop{\partial\kern -4pt\raisebox{.8pt}{$|$}}\wedge f-\Theta^{\rho}%
\wedge\left(  \theta_{\rho}\mathbin\lrcorner f\right)  \right)  \medskip
\nonumber\\
&  =\mbox{\boldmath$\partial$}\wedge\left(
\mathop{\partial\kern -4pt\raisebox{.8pt}{$|$}}\wedge f\right)  =\left(
\mathop{\partial\kern -4pt\raisebox{.8pt}{$|$}}\wedge-\Theta^{\rho}\wedge
i_{\rho}\right)  \left(  \mathop{\partial\kern -4pt\raisebox{.8pt}{$|$}}\wedge
f\right)  \medskip\nonumber\\
&  =\mathop{\partial\kern -4pt\raisebox{.8pt}{$|$}}\wedge\left(
\mathop{\partial\kern -4pt\raisebox{.8pt}{$|$}}\wedge f\right)  -\Theta^{\rho
}\wedge i_{\rho}\left(  \mathop{\partial\kern -4pt\raisebox{.8pt}{$|$}}\wedge
f\right)  =-\Theta^{\rho}\wedge i_{\rho}\left(
\mathop{\partial\kern -4pt\raisebox{.8pt}{$|$}}\wedge f\right)  \medskip
\label{f16}\\
&  =-\Theta^{\rho}\wedge\left(  \theta_{\rho}\mathbin\lrcorner\theta^{\alpha
}\mathbf{e}_{\alpha}\left(  f\right)  \right)  =-\Theta^{\rho}\delta_{\rho
}^{\alpha}\mathbf{e}_{\alpha}\left(  f\right)  =-\Theta^{\alpha}%
\mathbf{e}_{\alpha}\left(  f\right)  \medskip\nonumber\\
&  =-\frac{1}{2}T_{\rho\sigma}^{\alpha}\left(  \theta^{\rho}\wedge
\theta^{\sigma}\right)  \mathbf{e}_{\alpha}\left(  f\right)  .\nonumber
\end{align}

\end{description}

Finally, from the Eqs.(\ref{f14}) and (\ref{f16}) we get that
$\mbox{\boldmath$\partial$}^{2}f$ is given by%
\begin{equation}
\mbox{\boldmath$\partial$}^{2}f=g^{\beta\alpha}\nabla_{\mathbf{e}_{\beta}%
}\nabla_{\mathbf{e}_{\alpha}}f-g^{\beta\rho}\overset{\circ}{\Gamma}_{\beta
\rho}^{\alpha}\mathbf{e}_{\alpha}\left(  f\right)  +T_{\alpha\beta}^{\alpha
}\mathbf{e}^{\beta}\left(  f\right)  -\frac{1}{2}T_{\rho\sigma}^{\alpha
}\left(  \theta^{\rho}\wedge\theta^{\sigma}\right)  \mathbf{e}_{\alpha}\left(
f\right)  . \label{f17}%
\end{equation}
We now define
\[
T_{\alpha\beta}^{\alpha}\equiv Q_{\beta},
\]
then the Eq.(\ref{f17}) can be written as%
\begin{equation}
\mbox{\boldmath$\partial$}^{2}f=g^{\beta\alpha}\nabla_{\mathbf{e}_{\beta}%
}\nabla_{\mathbf{e}_{\alpha}}f+g^{\beta\rho}\overset{\circ}{\Gamma}_{\beta
\rho}^{\alpha}\mathbf{e}_{\alpha}\left(  f\right)  +Q_{\beta}\mathbf{e}%
^{\beta}\left(  f\right)  -\frac{1}{2}T_{\rho\sigma}^{\alpha}\left(
\theta^{\rho}\wedge\theta^{\sigma}\right)  \mathbf{e}_{\alpha}\left(
f\right)  . \label{f18}%
\end{equation}

In general, we can calculate the square of the Dirac operator directly from
the definition, i.e., without using of the standard Dirac operator. Indeed, we
can write%

\[%
\begin{array}
[c]{l}%
\mbox{\boldmath$\partial$}^{2} =\left(  \theta^{\beta}\nabla_{\mathbf{e}%
_{\beta}}\right)  \left(  \theta^{\rho}\nabla_{\rho_{\rho}}\right)  \medskip\\
=\theta^{\beta}\left[  \theta^{\rho}\left(  \nabla_{\mathbf{e}_{\beta}}%
\nabla_{\mathbf{e}_{\rho}}\right)  +\left(  \nabla_{\mathbf{e}_{\beta}}%
\theta^{\rho}\right)  \nabla_{\mathbf{e}_{\rho}}\right]  \medskip\\
=\theta^{\beta}\mathbin\lrcorner\left[  \theta^{\rho}\left(  \nabla
_{\mathbf{e}_{\beta}}\nabla_{\mathbf{e}_{\rho}}\right)  +\left(
\nabla_{\mathbf{e}_{\beta}}\theta^{\rho}\right)  \nabla_{\mathbf{e}_{\rho}%
}\right]  +\theta^{\beta}\wedge\left[  \theta^{\rho}\left(  \nabla
_{\mathbf{e}_{\beta}}\nabla_{\mathbf{e}_{\rho}}\right)  +\left(
\nabla_{\mathbf{e}_{\beta}}\theta^{\rho}\right)  \nabla_{\mathbf{e}_{\rho}%
}\right]  \medskip\\
=\theta^{\beta}\cdot\theta^{\rho}\left(  \nabla_{\mathbf{e}_{\beta}}%
\nabla_{\mathbf{e}_{\rho}}\right)  +\theta^{\beta}\mathbin\lrcorner\left(
-\Gamma_{\beta\alpha}^{\rho}\theta^{\alpha}\right)  \nabla_{\mathbf{e}_{\rho}%
}\medskip\\
+\theta^{\beta}\wedge\theta^{\rho}\left(  \nabla_{\mathbf{e}_{\beta}}%
\nabla_{\mathbf{e}_{\rho}}\right)  +\theta^{\beta}\wedge\left(  -\Gamma
_{\beta\alpha}^{\rho}\theta^{\alpha}\right)  \nabla_{\mathbf{e}_{\rho}%
}\medskip\\
=g^{\beta\rho}\left[  \nabla_{\mathbf{e}_{\beta}}\nabla_{\mathbf{e}_{\rho}%
}-\Gamma_{\beta\rho}^{\alpha}\nabla_{\mathbf{e}_{\alpha}}\right]
+\theta^{\beta}\wedge\theta^{\rho}\left[  \nabla_{\mathbf{e}_{\beta}}%
\nabla_{\mathbf{e}_{\rho}}-\Gamma_{\beta\rho}^{\alpha}\nabla_{\mathbf{e}%
_{\alpha}}\right]  .
\end{array}
\]

So, we have%
\begin{equation}
\mbox{\boldmath$\partial$}^{2} =g^{\beta\rho}\left[  \nabla_{\mathbf{e}%
_{\beta}}\nabla_{\mathbf{e}_{\rho}}-\Gamma_{\beta\rho}^{\alpha}\nabla
_{\mathbf{e}_{\alpha}}\right]  +\theta^{\beta}\wedge\theta^{\rho}\left[
\nabla_{\mathbf{e}_{\beta}}\nabla_{\mathbf{e}_{\rho}}-\Gamma_{\beta\rho
}^{\alpha}\nabla_{\mathbf{e}_{\alpha}}\right]  , \label{f19}%
\end{equation}

where we wrote
\[
\theta^{\beta}\mathbin\lrcorner\left(  -\Gamma_{\beta\alpha}^{\rho}%
\theta^{\alpha}\right)  \nabla_{\mathbf{e}_{\rho}}=-\theta^{\beta}\cdot
\theta^{\alpha}\Gamma_{\beta\alpha}^{\rho}\nabla_{\mathbf{e}_{\rho}}%
=-g^{\beta\rho}\Gamma_{\beta\alpha}^{\rho}\nabla_{\mathbf{e}_{\alpha}}%
\]
and%
\[
\theta^{\beta}\wedge\left(  -\Gamma_{\beta\alpha}^{\rho}\theta^{\alpha
}\right)  \nabla_{\mathbf{e}_{\rho}}=-\theta^{\beta}\wedge\theta^{\alpha
}\Gamma_{\beta\alpha}^{\rho}\nabla_{\mathbf{e}_{\rho}}=-\theta^{\beta}%
\wedge\theta^{\rho}\Gamma_{\beta\alpha}^{\rho}\nabla_{\mathbf{e}_{\alpha}}.
\]

On the other hand, the second term of the right hand side of the
Eq.(\ref{f19}), can be written as%
\begin{equation}%
\begin{array}
[c]{l}%
\theta^{\beta}\wedge\theta^{\rho}\left[  \nabla_{\mathbf{e}_{\beta}}%
\nabla_{\mathbf{e}_{\rho}}-\Gamma_{\beta\rho}^{\alpha}\nabla_{\mathbf{e}%
_{\alpha}}\right]  \medskip\\
=\frac{1}{2}\theta^{\beta}\wedge\theta^{\rho}\left[  \nabla_{\mathbf{e}%
_{\beta}}\nabla_{\mathbf{e}_{\rho}}-\Gamma_{\beta\rho}^{\alpha}\nabla
_{\mathbf{e}_{\alpha}}\right]  +\frac{1}{2}\theta^{\rho}\wedge\theta^{\beta
}\left[  \nabla_{\mathbf{e}_{\rho}}\nabla_{\mathbf{e}_{\beta}}-\Gamma
_{\rho\beta}^{\alpha}\nabla_{\mathbf{e}_{\alpha}}\right]  \medskip\\
=\frac{1}{2}\theta^{\beta}\wedge\theta^{\rho}\left[  \nabla_{\mathbf{e}%
_{\beta}}\nabla_{\mathbf{e}_{\rho}}-\nabla_{\mathbf{e}_{\rho}}\nabla
_{\mathbf{e}_{\beta}}-\left(  \Gamma_{\beta\rho}^{\alpha}-\Gamma_{\rho\beta
}^{\alpha}\right)  \nabla_{\mathbf{e}_{\alpha}}\right]  .
\end{array}
\label{f20}%
\end{equation}
So, from the Eqs.(\ref{f19}) and (\ref{f20}) we get%
\begin{equation}%
\begin{array}
[c]{ll}%
\mbox{\boldmath$\partial$}^{2} & =g^{\beta\rho}\left[  \nabla_{\mathbf{e}%
_{\beta}}\nabla_{\mathbf{e}_{\rho}}-\Gamma_{\beta\rho}^{\alpha}\nabla
_{\mathbf{e}_{\alpha}}\right]  \medskip\\
& +\frac{1}{2}\theta^{\beta}\wedge\theta^{\rho}\left[  \nabla_{\mathbf{e}%
_{\beta}}\nabla_{\mathbf{e}_{\rho}}-\nabla_{\mathbf{e}_{\rho}}\nabla
_{\mathbf{e}_{\beta}}-\left(  \Gamma_{\beta\rho}^{\alpha}-\Gamma_{\rho\beta
}^{\alpha}\right)  \nabla_{\mathbf{e}_{\alpha}}\right]  .
\end{array}
\label{f21}%
\end{equation}

Now, let $f\in\sec%
{\displaystyle\bigwedge\nolimits^{0}}
T^{\ast}M\hookrightarrow\sec$\texttt{ }$\mathcal{C}\ell\left(  M,\mathtt{g}%
\right)  $. Then, using Eq.(\ref{f21}) we can calculate
$\mbox{\boldmath$\partial$}^{2} f$ as follows.%
\begin{equation}%
\begin{array}
[c]{ll}%
\mbox{\boldmath$\partial$}^{2} f & =g^{\beta\rho}\left[  \nabla_{\mathbf{e}%
_{\beta}}\nabla_{\mathbf{e}_{\rho}}-\Gamma_{\beta\rho}^{\alpha}\nabla
_{\mathbf{e}_{\alpha}}\right]  f\medskip\\
& +\frac{1}{2}\theta^{\beta}\wedge\theta^{\rho}\left[  \nabla_{\mathbf{e}%
_{\beta}}\nabla_{\mathbf{e}_{\rho}}-\nabla_{\mathbf{e}_{\rho}}\nabla
_{\mathbf{e}_{\beta}}-\left(  \Gamma_{\beta\rho}^{\alpha}-\Gamma_{\rho\beta
}^{\alpha}\right)  \nabla_{\mathbf{e}_{\alpha}}\right]  f.
\end{array}
\label{f22}%
\end{equation}
On the other hand, observe that%
\[
\nabla_{\mathbf{e}_{\beta}}\nabla_{\mathbf{e}_{\rho}}f-\nabla_{\mathbf{e}%
_{\rho}}\nabla_{\mathbf{e}_{\beta}}f=\left[  \mathbf{e}_{\beta},\mathbf{e}%
_{\rho}\right]  f=c_{\beta\rho}^{\alpha}\mathbf{e}_{\alpha}\left(  f\right)
,
\]
and recalling that $T_{\beta\rho}^{\alpha}\equiv\Gamma_{\beta\rho}^{\alpha
}-\Gamma_{\rho\beta}^{\alpha}-c_{\beta\rho}^{\alpha}$, Eq.(\ref{f22}) can
be\textbf{ }written as%

\begin{equation}
\mbox{\boldmath$\partial$}^{2} f=g^{\beta\rho}\nabla_{\mathbf{e}_{\beta}%
}\nabla_{\mathbf{e}_{\rho}}f-g^{\beta\rho}\Gamma_{\beta\rho}^{\alpha
}\mathbf{e}_{\alpha}\left(  f\right)  -\frac{1}{2}\theta^{\beta}\wedge
\theta^{\rho}T_{\beta\rho}^{\alpha}\mathbf{e}_{\alpha}\left(  f\right)  .
\label{f23}%
\end{equation}
Note that, for the particular case of a coordinate basis, $\mathbf{e}_{\alpha
}=\frac{\partial}{\partial x^{\alpha}}$, the $c_{\beta\rho}^{\alpha}=0$ and we
have%
\[
T_{\beta\rho}^{\alpha}\equiv\Gamma_{\beta\rho}^{\alpha}-\Gamma_{\rho\beta
}^{\alpha}.
\]

Now, we must show the equivalence of the Eqs.(\ref{f23}) and (\ref{f17}). For
that, we use a well known relation between the covariant derivatives $D$ and
$\nabla$ saying that (see, e.g.,\cite{rodoliv2006})%
\begin{equation}
K_{\beta\rho}^{\alpha}=\Gamma_{\beta\rho}^{\alpha}-\mathring{\Gamma}%
_{\beta\rho}^{\alpha},\label{f24}%
\end{equation}
where $K_{\beta\rho}^{\alpha}$ is the so-called cotorsion tensor (see, e.g.,
\cite{rodoliv2006}), given by%
\begin{equation}
K_{\beta\rho}^{\alpha}=-\frac{1}{2}g^{\alpha\sigma}\left[  g_{\mu\beta}%
T_{\rho\sigma}^{\mu}+g_{\mu\rho}T_{\beta\sigma}^{\mu}-g_{\mu\sigma}%
T_{\beta\rho}^{\mu}\right]  .\label{f24a}%
\end{equation}
Comparing the Eq.(\ref{f17}) and Eq.(\ref{f23}), we must show that%
\[
g^{\beta\rho}\Gamma_{\beta\rho}^{\alpha}\mathbf{e}_{\alpha}\left(  f\right)
=g^{\beta\rho}\mathring{\Gamma}_{\beta\rho}^{\alpha}\mathbf{e}_{\alpha}\left(
f\right)  -T_{\alpha\beta}^{\alpha}\mathbf{e}^{\beta}\left(  f\right)  ,
\]
or
\begin{equation}
g^{\beta\rho}\Gamma_{\beta\rho}^{\alpha}\mathbf{e}_{\alpha}\left(  f\right)
=g^{\beta\rho}\mathring{\Gamma}_{\beta\rho}^{\alpha}\mathbf{e}_{\alpha}\left(
f\right)  -T_{\alpha\beta}^{\alpha}g^{\beta\delta}\mathbf{e}_{\delta}\left(
f\right)  .\label{f25}%
\end{equation}
From the Eq.(\ref{f24}), we see that
\begin{equation}
g^{\beta\rho}\Gamma_{\beta\rho}^{\alpha}\mathbf{e}_{\alpha}\left(  f\right)
=g^{\beta\rho}\mathring{\Gamma}_{\beta\rho}^{\alpha}\mathbf{e}_{\alpha}\left(
f\right)  +g^{\beta\rho}K_{\beta\rho}^{\alpha}\mathbf{e}_{\alpha}\left(
f\right)  .\label{f26}%
\end{equation}

The second term of the right side of Eq.(\ref{f26}) can be written as%
\begin{align}
K_{\beta\rho}^{\alpha}g^{\beta\rho} &  =-\frac{1}{2}g^{\alpha\sigma}\left[
g_{\mu\beta}T_{\rho\sigma}^{\mu}+g_{\mu\rho}T_{\beta\sigma}^{\mu}-g_{\mu
\sigma}T_{\beta\rho}^{\mu}\right]  g^{\beta\rho}\nonumber\\
&  =-T_{\rho\sigma}^{\rho}g^{\sigma\alpha}+\frac{1}{2}T_{\beta\mu}^{\alpha
}g^{\beta\mu}\nonumber\\
&  =-T_{\rho\sigma}^{\rho}g^{\sigma\alpha},\label{f27}%
\end{align}
where we used the fact that $T_{\beta\mu}^{\alpha}g^{\beta\mu}=0$. Finally,
inserting Eq.(\ref{f27}) into Eq.(\ref{f26}), we obtain
\begin{equation}
g^{\beta\rho}\Gamma_{\beta\rho}^{\alpha}\mathbf{e}_{\alpha}\left(  f\right)
=g^{\beta\rho}\mathring{\Gamma}_{\beta\rho}^{\alpha}\mathbf{e}_{\alpha}\left(
f\right)  -T_{\rho\sigma}^{\rho}g^{\sigma\alpha}\mathbf{e}_{\alpha}\left(
f\right)  ,\label{f28}%
\end{equation}
and Eq.(\ref{f25}) is proved. Our result is to be compared with the one in
\cite{Ra1}, which is unfortunately equivocated.

\subsection{Relation Between the Dirac Operators Associated with $D$ and
$\nabla$ for the Case of Null Strain Tensor}

Let us now write the square of the Dirac operator on a Riemann-Cartan space
$(M,\mathbf{g,}\nabla$) in terms of the square of standard Dirac operator
acting on a Riemannian (or Lorentzian) space(time) $(M,\mathbf{g,}D$), for the
case of a null strain tensor.

We start recalling the well known relation between the connection coefficients
(in an arbitrary basis) of a general Riemann-Cartan connection $\nabla$ and
$D$, which in a is given by
\begin{equation}
\Gamma_{\alpha\beta}^{\rho}=\mathring{\Gamma}_{\alpha\beta}^{\rho}+\frac{1}%
{2}T_{\alpha\beta}^{\rho}+\frac{1}{2}S_{\alpha\beta}^{\rho},\label{r1}%
\end{equation}
where as before $T_{\alpha\beta}^{\rho}$ are the components of the torsion
tensor and $S_{\alpha\beta}^{\rho}$ are the components of the strain tensor of
the connection (see details, e.g., in \cite{rodoliv2006}). For what follows we
are interested in the important case where $S_{\alpha\beta}^{\rho}=0$, in
which case Eq.(\ref{r1}) reduces to
\begin{equation}
\Gamma_{\alpha\beta}^{\rho}=\mathring{\Gamma}_{\alpha\beta}^{\rho}+\frac{1}%
{2}T_{\alpha\beta}^{\rho}.\label{r2}%
\end{equation}
Now, recalling that $D_{\mathbf{e}_{\beta}}\mathbf{e}_{\alpha}=\mathring
{\Gamma}_{\beta\alpha}^{\rho}\mathbf{e}_{\rho}$, and $\nabla_{\mathbf{e}%
_{\beta}}\mathbf{e}_{\alpha}=\Gamma_{\beta\alpha}^{\rho}\mathbf{e}_{\rho}$ we
have immediately%
\begin{equation}
\nabla_{\mathbf{e}_{\alpha}}\mathbf{e}_{\alpha}=D_{\mathbf{e}_{\alpha}%
}\mathbf{e}_{\beta}+\frac{1}{2}\tau\left(  \mathbf{e}_{\alpha},\mathbf{e}%
_{\beta}\right)  .\label{r3}%
\end{equation}
\medskip where we recall that for any $\mathbf{u,v}\in\sec TM,$ with
$\mathbf{u}=u^{\alpha}\mathbf{e}_{\alpha}$ and $\mathbf{v}=v^{\beta}%
\mathbf{e}_{\beta}$, the torsion operator is given by

\medskip%
\begin{equation}
\tau\left(  \mathbf{u},\mathbf{v}\right)  =\tau\left(  u^{\alpha}%
\mathbf{e}_{\alpha},v^{\beta}\mathbf{e}_{\beta}\right)  =u^{\alpha}v^{\beta
}\left[  \Gamma_{\alpha\beta}^{\rho}-\Gamma_{\beta\alpha}^{\rho}%
-c_{\alpha\beta}^{\rho}\right]  \mathbf{e}_{\rho}=u^{\alpha}v^{\beta}%
T_{\alpha\beta}^{\rho}\mathbf{e}_{\rho}, \label{r4}%
\end{equation}
where $c_{\alpha\beta}^{\rho}=\left[  \mathbf{e}_{\alpha},\mathbf{e}_{\beta
}\right]  .$ Using Eq.(\ref{r4}) we can calculate $\nabla_{\mathbf{u}%
}\mathbf{v},$ in terms of $D_{\mathbf{u}}\mathbf{v}$. We have:%

\begin{equation}
\nabla_{\mathbf{u}}\mathbf{v}=D_{\mathbf{u}}\mathbf{v}+\frac{1}{2}\tau\left(
\mathbf{u,v}\right)  . \label{r5}%
\end{equation}

On the other hand, recalling Eqs. (\ref{f1}) and (\ref{f3}),
\[%
\bpartial
=\theta^{\alpha}D_{\mathbf{e}_{\alpha}}\qquad\text{and\qquad}%
\mbox{\boldmath$\partial$}=\theta^{\alpha}\nabla_{\mathbf{e}_{\alpha}}%
\]
we have for $A\in\sec TM$
\begin{equation}
\mbox{\boldmath$\partial$} A=%
\bpartial
A+\frac{1}{2}\theta^{\alpha}\tau\left(  \mathbf{e}_{\alpha},A\right)  .
\label{r6}%
\end{equation}

\begin{remark}
Notice that if $\tau\left(  \mathbf{u,v}\right)  =0,$ then
$\mbox{\boldmath$\partial$}=%
\bpartial
.$
\end{remark}

Now, from Eq.(\ref{r5}) we can exhibit a relation between
$\mbox{\boldmath$\partial$}^{2}$ and $%
\bpartial
^{2}$, a task that is made easier if we use of a pair of dual orthonormal
basis $\{\mathbf{e}_{\mathbf{a}}\}$ and $\{\theta^{\mathbf{b}}\}$ for $TU$ and
$T^{\ast}U$ ($U\subset M$). Indeed, from Eq.(\ref{f19}) we immediately have
\begin{align}
\mbox{\boldmath$\partial$}^{2}  &  =\mbox{\boldmath$\partial$}\cdot
\mbox{\boldmath$\partial$}+\mbox{\boldmath$\partial$}\wedge
\mbox{\boldmath$\partial$}\nonumber\\
&  =\eta^{\mathbf{ab}}\left[  \nabla_{\mathbf{e}_{\mathbf{a}}}\nabla
_{\mathbf{e}_{\mathbf{b}}}-\Gamma_{\mathbf{ab}}^{\mathbf{c}}\nabla
_{\mathbf{e}_{\mathbf{c}}}\right]  +\theta^{\mathbf{a}}\wedge\theta
^{\mathbf{b}}\left[  \nabla_{\mathbf{e}_{\mathbf{a}}}\nabla_{\mathbf{e}%
_{\mathbf{b}}}-\Gamma_{\mathbf{ab}}^{\mathbf{c}}\nabla_{\mathbf{e}%
_{\mathbf{c}}}\right]  , \label{r7}%
\end{align}
where $\eta^{\mathbf{ab}}=\eta_{\mathbf{ab}}=\mathrm{diag}(1,-1,-1,-1).$

In order to write Eq.(\ref{r7}) in terms of $D$, first we calculate
$\nabla_{\mathbf{e}_{\beta}}\nabla_{\mathbf{e}_{\rho}}A$ where $A\in\sec TM$,
then from Eq.(\ref{r5}) we have%
\begin{equation}
\nabla_{\mathbf{e}_{\mathbf{a}}}\nabla_{\mathbf{e}_{\mathbf{b}}}%
A=D_{\mathbf{e}_{\mathbf{a}}}D_{\mathbf{e}_{\mathbf{b}}}A+\frac{1}{2}%
\tau\left(  \mathbf{e}_{\mathbf{a}},D_{\mathbf{e}_{\mathbf{b}}}A\right)
+\frac{1}{2}D_{\mathbf{e}_{\mathbf{a}}}\tau\left(  \mathbf{e}_{\mathbf{b}%
},A\right)  +\frac{1}{4}\tau\left(  \mathbf{e}_{\mathbf{a}},\tau\left(
\mathbf{e}_{\mathbf{b}},A\right)  \right)  . \label{p1}%
\end{equation}

On the other hand%
\begin{equation}
\Gamma_{\mathbf{ab}}^{\mathbf{c}}\nabla_{\mathbf{e}_{\mathbf{c}}}A=\left(
\mathring{\Gamma}_{\mathbf{ab}}^{\mathbf{c}}+\frac{1}{2}T_{\mathbf{ab}%
}^{\mathbf{c}}\right)  \left(  D_{\mathbf{e}_{\mathbf{c}}}A+\frac{1}{2}%
\tau\left(  \mathbf{e}_{\mathbf{c}},A\right)  \right)  ,\label{p2}%
\end{equation}
and from Eq.(\ref{p1}) and Eq.(\ref{p2}) we can write%
\[%
\begin{array}
[c]{ll}%
\mbox{\boldmath$\partial$}\cdot\mbox{\boldmath$\partial$}A & =\eta
^{\mathbf{ab}}\left\{  D_{\mathbf{e}_{\mathbf{a}}}D_{\mathbf{e}_{\mathbf{b}}%
}A+\frac{1}{2}\tau\left(  \mathbf{e}_{\mathbf{a}},D_{\mathbf{e}_{\mathbf{b}}%
}A\right)  +\frac{1}{2}D_{\mathbf{e}_{\mathbf{a}}}\tau\left(  \mathbf{e}%
_{\mathbf{b}},A\right)  \right.  \medskip\\
& \left.  +\frac{1}{4}\tau\left(  \mathbf{e}_{\mathbf{a}},\tau\left(
\mathbf{e}_{\mathbf{b}},A\right)  \right)  -\left(  \mathring{\Gamma
}_{\mathbf{ab}}^{\mathbf{c}}+\frac{1}{2}T_{\mathbf{ab}}^{\mathbf{c}}\right)
\left(  D_{\mathbf{e}_{\mathbf{c}}}A+\frac{1}{2}\tau\left(  \mathbf{e}%
_{\mathbf{c}},A\right)  \right)  \right\}  \medskip\\
& =\eta^{\mathbf{ab}}\left\{  D_{\mathbf{e}_{\mathbf{a}}}D_{\mathbf{e}%
_{\mathbf{b}}}A-\mathring{\Gamma}_{\mathbf{ab}}^{\mathbf{c}}D_{\mathbf{e}%
_{\mathbf{c}}}A+\frac{1}{2}\tau\left(  \mathbf{e}_{\mathbf{a}},D_{\mathbf{e}%
_{\mathbf{b}}}A\right)  +\frac{1}{2}D_{\mathbf{e}_{\mathbf{a}}}\tau\left(
\mathbf{e}_{\mathbf{b}},A\right)  \right.  \medskip\\
& \left.  +\frac{1}{4}\tau\left(  \mathbf{e}_{\mathbf{a}},\tau\left(
\mathbf{e}_{\mathbf{b}},A\right)  \right)  -\frac{1}{2}T_{\mathbf{ab}%
}^{\mathbf{c}}D_{\mathbf{e}_{\mathbf{c}}}A-\frac{1}{2}\mathring{\Gamma
}_{\mathbf{ab}}^{\mathbf{c}}\tau\left(  \mathbf{e}_{\mathbf{c}},A\right)
-\frac{1}{4}T_{\mathbf{ab}}^{\mathbf{c}}\tau\left(  \mathbf{e}_{\mathbf{c}%
},A\right)  \right\}  ,
\end{array}
\]
or%
\begin{equation}%
\begin{array}
[c]{ll}%
\mbox{\boldmath$\partial$}\cdot\mbox{\boldmath$\partial$}A & =\eta
^{\mathbf{ab}}\left\{
\bpartial
\cdot%
\bpartial
A+\frac{1}{2}\tau\left(  \mathbf{e}_{\mathbf{a}},D_{\mathbf{e}_{\mathbf{b}}%
}A\right)  +\frac{1}{2}D_{\mathbf{e}_{\mathbf{a}}}\tau\left(  \mathbf{e}%
_{\mathbf{b}},A\right)  -\frac{1}{2}T_{\mathbf{ab}}^{\mathbf{c}}%
D_{\mathbf{e}_{\mathbf{c}}}A\right.  \medskip\\
& \left.  -\frac{1}{2}\mathring{\Gamma}_{\mathbf{ab}}^{\mathbf{c}}\tau\left(
\mathbf{e}_{\mathbf{c}},A\right)  +\frac{1}{4}\tau\left(  \mathbf{e}%
_{\mathbf{a}},\tau\left(  \mathbf{e}_{\mathbf{b}},A\right)  \right)  -\frac
{1}{4}T_{\mathbf{ab}}^{\mathbf{c}}\tau\left(  \mathbf{e}_{\mathbf{c}%
},A\right)  \right\}  .
\end{array}
\label{p3}%
\end{equation}

Thus, using Eq.(\ref{p3}) and recalling that $\mbox{\boldmath$\partial$}^{2}%
=\mbox{\boldmath$\partial$}\cdot
\mbox{\boldmath$\partial$}+\mbox{\boldmath$\partial$}\wedge
\mbox{\boldmath$\partial$}$ we get
\begin{equation}
\mbox{\boldmath$\partial$}^{2}A=\left(
\bpartial
\right)  ^{2}A+g^{\mathbf{ab}}S_{1}A+\theta^{\mathbf{a}}\wedge\theta
^{\mathbf{b}}S_{1}A,\label{p4}%
\end{equation}
where%
\begin{equation}%
\begin{array}
[c]{ll}%
S_{1}A & =\frac{1}{2}\tau\left(  \mathbf{e}_{\mathbf{a}},D_{\mathbf{e}%
_{\mathbf{b}}}A\right)  +\frac{1}{2}D_{\mathbf{e}_{\mathbf{a}}}\tau\left(
\mathbf{e}_{\mathbf{b}},A\right)  -\frac{1}{2}T_{\mathbf{ab}}^{\mathbf{c}%
}D_{\mathbf{e}_{\mathbf{c}}}A\medskip\\
& -\frac{1}{2}\mathring{\Gamma}_{\mathbf{ab}}^{\mathbf{c}}\tau\left(
\mathbf{e}_{\mathbf{c}},A\right)  +\frac{1}{4}\tau\left(  \mathbf{e}%
_{\mathbf{a}},\tau\left(  \mathbf{e}_{\mathbf{b}},A\right)  \right)  -\frac
{1}{4}T_{\mathbf{ab}}^{\mathbf{c}}\tau\left(  \mathbf{e}_{\mathbf{c}%
},A\right)  .
\end{array}
\label{p5}%
\end{equation}

\section{Spinor Bundles and Spinor Fields}

In what follows we assume that ($M,\mathbf{g})$ is a $4$-dimensional spin
manifold representing a \textit{spacetime} \cite{LaMi, rodoliv2006}. We start
by recalling the noticeable results of \cite{moro,rodoliv2006} on the
possibility of defining "unit sections" on various different vector bundles
associated with the principal bundle $P_{\mathrm{Spin}_{1,3}^{e}}\left(
M\right)  $, the \textit{covering} of $P_{\mathrm{SO}_{1,3}^{e}}\left(
M\right)  $, the orthonormal frame bundle.

Let
\[
\Phi_{i}:\pi^{-1}\left(  U_{i}\right)  \rightarrow U_{i}\times\mathrm{Spin}%
_{1,3}^{e},\qquad\Phi_{j}:\pi^{-1}\left(  U_{j}\right)  \rightarrow
U_{j}\times\mathrm{Spin}_{1,3}^{e},
\]
be two local trivializations for $P_{\mathrm{Spin}_{1,3}^{e}}\left(  M\right)
$, with
\[
\Phi_{i}\left(  u\right)  =\left(  \pi\left(  u\right)  =x,\phi_{i,x}\left(
u\right)  \right)  ,\qquad\Phi_{j}\left(  u\right)  =\left(  \pi\left(
u\right)  =x,\phi_{j,x}\left(  u\right)  \right)  .
\]
Recall that the transition function $\mathtt{h}_{ij}:U_{i}\cap U_{j}%
\rightarrow\mathrm{Spin}_{1,3}^{e}$ is then given by%
\[
\mathtt{h}_{ij}\left(  x\right)  =\phi_{i,x}\circ\phi_{j,x}^{-1},
\]
which does not depend on $u.$

\begin{proposition}
$\mathcal{C}\ell(M,\mathtt{g})$ has a naturally defined global unit
section.\emph{\footnote{Recall that in a spin manifold the Clifford bundle
$\mathcal{C}\ell(M,\mathtt{g})$ can also be an assocaited vector bundle to the
principal bundle $P_{\mathrm{Spin}_{1,3}^{e}}$, i.e., $\mathcal{C}%
\ell_{\mathrm{Spin}_{1,3}^{e}}^{\ell}(M,\mathtt{g})=\mathbf{P}_{\mathrm{Spin}%
_{1,3}^{e}}\times_{Ad}\mathbb{R}_{1,3}$. See details in, e.g., \cite{LaMi,
rodoliv2006}.}}
\end{proposition}

\begin{proof}
For the associated bundle $\mathcal{C}\ell(M,\mathtt{g}),$ the transition
functions corresponding to local trivializations%
\begin{equation}
\Psi_{i}:\pi_{c}^{-1}\left(  U_{i}\right)  \rightarrow U_{i}\times
\mathbb{R}_{1,3},\qquad\Psi_{j}:\pi_{c}^{-1}\left(  U_{j}\right)  \rightarrow
U_{j}\times\mathbb{R}_{1,3} \label{s1}%
\end{equation}
are given by $h_{ij}\left(  x\right)  =Ad_{\mathtt{h}_{ij}\left(  x\right)
}.$ Define the local sections%
\begin{equation}
\mathbf{1}_{i}\left(  x\right)  =\Psi_{i}^{-1}\left(  x,1\right)
,\qquad\mathbf{1}_{j}\left(  x\right)  =\Psi_{j}^{-1}\left(  x,1\right)  ,
\label{s2}%
\end{equation}
where $1$ is the unit element of $\mathbb{R}_{1,3}$ (the \textit{spacetime
algebra}, see \emph{\cite{rodoliv2006}})$.$ Since
\[
h_{ij}\left(  x\right)  \cdot1=Ad_{\mathtt{h}_{ij}\left(  x\right)  }\left(
1\right)  =\mathtt{h}_{ij}\left(  x\right)  1\mathtt{h}_{ij}\left(  x\right)
^{-1}=1,
\]
we see that the expressions above uniquely define a global section
$\mathbf{1}\in\mathcal{C}\ell(M,$\texttt{g}$)$ with $\left.  \mathbf{1}%
\right\vert _{U_{i}}=\mathbf{1}_{i}.$ This proves the proposition.
\end{proof}

\begin{definition}
The left real spin-Clifford bundle of $M$ is the vector bundle
\begin{equation}
\mathcal{C}\ell_{\mathrm{Spin}_{1,3}^{e}}^{\ell}(M,\mathtt{g}%
)=P_{\mathrm{Spin}_{1,3}^{e}}\times_{l}\mathbb{R}_{1,3} \label{ss1}%
\end{equation}
where $l$ is the representation of $\mathrm{Spin}_{1,3}^{e}$ on $\mathbb{R}%
_{1,3}$ given by $l\left(  a\right)  x=ax.$ Sections of $\mathcal{C}%
\ell_{\mathrm{Spin}_{1,3}^{e}}^{\ell}(M,\mathtt{g})$ are called left
spin-Clifford fields. In a similar way a right spin-Clifford bundle
$\mathcal{C}\ell_{\mathrm{Spin}_{1,3}^{e}}^{r}(M,\mathtt{g})$ is defined, see
\emph{\cite{moro,rodoliv2006}}.
\end{definition}

\begin{remark}
It is clear that the above proposition can be immediately generalized for the
Clifford bundle $\mathcal{C}\ell_{p,q}(M,\mathtt{g}),$ of any $n$-dimensional
manifold endowed with a metric of arbitrary signature $\left(  p,q\right)  $
(where $n=p+q).$ Now, we observe also that the left (and also the right)
spin-Clifford bundle can be generalized in an obvious way for any spin
manifold of arbitrary finite dimension $n=p+q,$ with a metric of arbitrary
signature $\left(  p,q\right)  .$ However, another important difference
between $\mathcal{C}\ell_{p,q}(M,$\texttt{g}$)$ and $\mathcal{C}%
\ell_{\mathrm{Spin}_{p,q}^{\mathbf{e}}}^{\ell}(M,\mathtt{g})$ or
\emph{(}$\mathcal{C}\ell_{\mathrm{Spin}_{p,q}^{\mathbf{e}}}^{r}(M,\mathtt{g}%
)$\emph{) }is that these latter bundles only admit a global unit section if
they are trivial.
\end{remark}

\begin{proposition}
There exists an unit section on $\mathcal{C}\ell_{\mathrm{Spin}_{p,q}%
^{\mathbf{e}}}^{r}(M,\mathtt{g})$ \emph{(}and also on $\mathcal{C}%
\ell_{\mathrm{Spin}_{p,q}^{\mathbf{e}}}^{\ell}(M,\mathtt{g})\emph{)}$ if, and
only if, $P_{\mathrm{Spin}_{p,q}^{\mathbf{e}}}\left(  M\right)  $ is trivial.
\end{proposition}

\begin{proof}
We show the necessity for the case of $\mathcal{C}\ell_{\mathrm{Spin}%
_{p,q}^{\mathbf{e}}}^{r}(M,\mathtt{g})$,\footnote{The proof for the case of
$\mathcal{C}\ell_{\mathrm{Spin}_{p,q}^{e}}^{\ell}(M,\mathtt{g})$ is
analogous.} the sufficiency is trivial. For $\mathcal{C}\ell_{\mathrm{Spin}%
_{p,q}^{\mathbf{e}}}^{r}(M,\mathtt{g})$\ the transition functions,
corresponding to local trivializations
\begin{equation}
\zeta_{i}:\pi_{sc}^{-1}\left(  U_{i}\right)  \rightarrow U_{i}\times
\mathbb{R}_{p,q},\qquad\zeta_{j}:\pi_{sc}^{-1}\left(  U_{j}\right)
\rightarrow U_{j}\times\mathbb{R}_{p,q} \label{s3}%
\end{equation}
are given by $k_{ij}\left(  x\right)  =R_{\mathtt{h}_{ij}}{}\left(  x\right)
,$ with $R_{a}:\mathbb{R}_{p},_{q}\rightarrow\mathbb{R}_{p},_{q},$
$x\longmapsto xa^{-1}.$ Let $1$ be the unit element of $\mathbb{R}_{p},_{q}$.
An unit section in $\mathcal{C}\ell_{\mathrm{Spin}_{p,q}^{\mathbf{e}}}%
^{r}(M,\mathtt{g})$--- if it exists--- is written in terms of these two
trivializations as%
\begin{equation}
\mathbf{1}_{i}\left(  x\right)  =\zeta_{i}^{-1}\left(  x,1\right)
,\qquad\mathbf{1}_{j}\left(  x\right)  =\zeta_{j}^{-1}\left(  x,1\right)  ,
\label{s4}%
\end{equation}
and we must have $\mathbf{1}_{i}\left(  x\right)  =\mathbf{1}_{j}\left(
x\right)  \forall x\in U_{i}\cap U_{j}.$ As $\zeta_{i}\left(  \mathbf{1}%
_{i}\left(  x\right)  \right)  =\left(  x,1\right)  =\zeta_{j}\left(
\mathbf{1}_{j}\left(  x\right)  \right)  ,$ we have $\mathbf{1}_{i}\left(
x\right)  =\mathbf{1}_{j}\left(  x\right)  \Leftrightarrow1=k_{ij}\left(
x\right)  \cdot1\Leftrightarrow1=k_{ij}\left(  x\right)  \Leftrightarrow
\mathtt{h}_{ij}\left(  x\right)  =1.$ This proves the proposition.
\end{proof}

We now, recall without proof a theorem (see Geroch \cite{Ge1}) that is crucial
for these theories.

\begin{theorem}
For a 4-dimensional Lorentzian manifold $\left(  M,\mathtt{g}\right)  ,$ a
spin structure exists if and only if $P_{\mathrm{SO}_{1,3}^{e}}\left(
M\right)  $ is a trivial bundle.
\end{theorem}

Recall that a principal bundle is trivial, if and only if, it admits a global
section. Therefore, Geroch's result says that a (non-compact) spacetime admits
a spin structure, if and only if, it admits a (globally defined) Lorentz
frame. In fact, it is possible to replace $P_{\mathrm{SO}_{1,3}^{e}}\left(
M\right)  $ by $P_{\mathrm{Spin}_{1,3}^{e}}\left(  M\right)  $ in the above
theorem. In this way, when a (non-compact) spacetime admits a spin structure,
the bundle $P_{\mathrm{Spin}_{1,3}^{e}}\left(  M\right)  $ is trivial and,
therefore, every bundle associated with it is trivial. For general spin
manifolds, the bundle $P_{\mathrm{Spin}_{p,q}^{e}}\left(  M\right)  $ is not
necessarily trivial for arbitrary $\left(  p,q\right)  ,$ but Geroch's theorem
warrants that, for the special case $\left(  p,q\right)  =\left(  1,3\right)
$ with $M$ non-compact, $P_{\mathrm{Spin}_{1,3}^{e}}\left(  M\right)  $ is
trivial. Then the above proposition implies that $\mathcal{C}\ell
_{\mathrm{Spin}_{1,3}^{e}}^{r}(M,\mathtt{g})$ and also on $\mathcal{C}%
\ell_{\mathrm{Spin}_{1,3}^{e}}^{\ell}(M,\mathtt{g})$ have global "unit
section". It is most important to note, however, that each different choice of
a (global) trivialization $\zeta_{i}$ on $\mathcal{C}\ell_{\mathrm{Spin}%
_{1,3}^{e}}^{r}(M,\mathtt{g})$ (respectively $\mathcal{C}\ell_{\mathrm{Spin}%
_{1,3}^{e}}^{\ell}(M,\mathtt{g}))$ induces a different global unit section
$\mathbf{1}_{i}^{r}$ (respectively $\mathbf{1}_{i}^{\ell}).$ Therefore, even
in this case there is no canonical unit section on $\mathcal{C}\ell
_{\mathrm{Spin}_{p,q}^{\mathbf{e}}}^{r}(M,\mathtt{g})$ (respectively on
$\mathcal{C}\ell_{\mathrm{Spin}_{1,3}^{e}}^{\ell}(M,\mathtt{g})).$

Then, when a (non-compact) spacetime $M$ is a spin manifold, the bundle
$P_{\mathrm{Spin}_{1,3}^{e}}\left(  M\right)  $ admits global sections. With
this in mind, let us fix a \textit{spin frame} $%
\mbox{\boldmath{$\Xi$}}%
$ and its dual spin \textit{coframe} $\Xi$\textit{ }for $M.$ This induces a
global trivialization for $P_{\mathrm{Spin}_{1,3}^{e}}\left(  M\right)  $ and
of course of $P_{\mathrm{Spin}_{1,3}^{e}}\left(  M\right)  .$ We the
trivialization of $P_{\mathrm{Spin}_{1,3}^{e}}\left(  M\right)  $ by%
\[
\Phi_{\Xi}:P_{\mathrm{Spin}_{1,3}^{e}}\left(  M\right)  \rightarrow
M\times\mathrm{Spin}_{1,3}^{e},
\]
with $\Phi_{\Xi}^{-1}\left(  x,1\right)  =\Xi\left(  x\right)  .$ We recall
that a spin coframe $\Xi\in\sec P_{\mathrm{Spin}_{1,3}^{e}}\left(  M\right)  $
can also be used to induced a certain fiducial global section on the various
vector bundles associated with $P_{\mathrm{Spin}_{1,3}^{e}}\left(  M\right)
:$

\begin{description}
\item[i) $\mathcal{C}\ell(M,$\texttt{g}$)$] 

\item Let $\left\{  \mathbf{E}^{\mathbf{a}}\right\}  $ be a fixed orthonormal
basis of $\mathbb{R}^{1,3}\hookrightarrow\mathbb{R}_{1,3}$ (which can be
thought of as the canonical basis of $\mathbb{R}^{1,3}$, the Minkowski
\textit{vector} space). We define basis sections in $\mathcal{C}\ell
(M,$\texttt{g}$)=P_{\mathrm{Spin}_{1,3}^{e}}\left(  M\right)  \times
_{Ad}\mathbb{R}_{1,3}$ by $\theta_{\mathbf{a}}\left(  x\right)  =\left[
\left(  \Xi\left(  x\right)  ,\mathbf{E}_{\mathbf{a}}\right)  \right]  .$ Of
course, this induces a multiform basis $\left\{  \theta_{I}\left(  x\right)
\right\}  $ for each $x\in M.$ Note that a more precise notation for
$\theta_{\mathbf{a}}$ would be, for instance, $\theta_{\mathbf{a}}^{\left(
\Xi\right)  }$.

\item[ii) $\mathcal{C}\ell_{\mathrm{Spin}_{1,3}^{e}}^{\ell}(M,\mathtt{g})$] 

\item Let $\mathbf{1}_{\Xi}^{\ell}\in\mathcal{C}\ell_{\mathrm{Spin}_{1,3}^{e}%
}^{\ell}(M,\mathtt{g})$ be defined by $\mathbf{1}_{\Xi}^{\ell}\left(
x\right)  \in\left[  \left(  \Xi\left(  x\right)  ,1\right)  \right]  .$ Then
the natural right action of $\mathbb{R}_{1,3}$ on $\mathcal{C}\ell
_{\mathrm{Spin}_{1,3}^{e}}^{\ell}(M,\mathtt{g})$ leads to $\mathbf{1}_{\Xi
}^{\ell}\left(  x\right)  a\in\left[  \left(  \Xi\left(  x\right)  ,a\right)
\right]  $ for all $a\in\mathbb{R}_{1,3}$.

\item[iii) $\mathcal{C}\ell_{\mathrm{Spin}_{1,3}^{e}}^{r}(M,\mathtt{g})$] 

\item Let $\mathbf{1}_{\Xi}^{r}\in\mathcal{C}\ell_{\mathrm{Spin}_{1,3}^{e}%
}^{r}(M,\mathtt{g})$ be defined by $\mathbf{1}_{\Xi}^{r}\left(  x\right)
\in\left[  \left(  \Xi\left(  x\right)  ,1\right)  \right]  .$ Then the
natural left action of $\mathbb{R}_{1,3}$ on $\mathcal{C}\ell_{\mathrm{Spin}%
_{1,3}^{e}}^{r}(M,\mathtt{g})$ leads to $\mathbf{1}_{\Xi}^{r}\left(  x\right)
a\in\left[  \left(  \Xi\left(  x\right)  ,a\right)  \right]  $ for all
$a\in\mathbb{R}_{1,3}$.
\end{description}

We now introduce without proof, some propositions which are crucial for our
calculations (for details, see \cite{rodoliv2006}).

\begin{proposition}
\begin{description}
\item[a)] $\mathbf{E}_{\mathbf{a}}=\mathbf{1}_{\Xi}^{r}\left(  x\right)
\theta_{\mathbf{a}}\left(  x\right)  \mathbf{1}_{\Xi}^{l}\left(  x\right)
,\quad\forall x\in M,$

\item[b)] $\mathbf{1}_{\Xi}^{l}\left(  x\right)  \mathbf{1}_{\Xi}^{r}\left(
x\right)  =1\in\mathcal{C}\ell(M,\mathtt{g}),$

\item[c)] $\mathbf{1}_{\Xi}^{r}\left(  x\right)  \mathbf{1}_{\Xi}^{l}\left(
x\right)  =1\in\mathbb{R}_{1,3}.$
\end{description}
\end{proposition}

\begin{proposition}
Let $\Xi,\Xi^{\prime}\in\sec P_{\mathrm{Spin}_{1,3}^{e}}\left(  M\right)  $ be
two spin coframes related by $\Xi^{\prime}=\Xi u,$ where $u:M\rightarrow
\mathrm{Spin}_{1,3}^{e}.$ Then%
\[%
\begin{array}
[c]{lll}%
a) & \theta_{\mathbf{a}}^{\prime}= & U\theta_{\mathbf{a}}U^{-1}\\
b) & \mathbf{1}_{\Xi^{\prime}}^{l}= & \mathbf{1}_{\Xi}^{l}u=U\mathbf{1}_{\Xi
}^{l},\\
c) & \mathbf{1}_{\Xi^{\prime}}^{r}= & u^{-1}\mathbf{1}_{\Xi}^{r}%
=\mathbf{1}_{\Xi}^{r}U^{-1}%
\end{array}
\]
where $U\in\sec\mathcal{C}\ell(M,\mathtt{g})$ is the Clifford field associated
with $u$ by $U\left(  x\right)  =\left[  \left(  \Xi\left(  x\right)
,u\left(  x\right)  \right)  \right]  .$ Also, b) and c), $u$ and $u^{-1}$
respectively act on $\mathbf{1}_{\Xi}^{l}\in\mathcal{C}\ell_{\mathrm{Spin}%
_{1,3}^{e}}^{\ell}(M,\mathtt{g})$ and $\mathbf{1}_{\Xi}^{r}\in\mathcal{C}%
\ell_{\mathrm{Spin}_{1,3}^{e}}^{r}(M,\mathtt{g}).$
\end{proposition}

\subsection{Covariant Derivatives of Clifford and Spinor Fields}

Since the Clifford bundle of differential forms is $\mathcal{C}\ell
(M,\mathtt{g})={\LARGE \tau}M/J_{\mathtt{g}},$ it is clear that any linear
connection $\nabla$ on the tensor bundle of \textit{covariant} tensors
${\LARGE \tau}M$ which is metric compatible $\left(  \nabla\mathbf{g}%
=\nabla\mathtt{g}\mathbf{=}0\right)  $ passes to the quotient ${\LARGE \tau
}M/J_{\mathtt{g}},$ and thus define an algebra bundle connection \cite{Cr}. On
the other hand, the spinor bundle $\mathcal{C}\ell_{\mathrm{Spin}_{1,3}^{e}%
}^{\ell}(M,\mathtt{g})$ and $\mathcal{C}\ell_{\mathrm{Spin}_{1,3}^{e}}%
^{r}(M,\mathtt{g})$ are vector bundles, thus as in the case of Clifford fields
we can use the general theory of covariant derivative operators on associated
vector bundles to obtain formulas for the covariant derivatives on sections of
these bundles. Given $\Psi\in\mathcal{C}\ell_{\mathrm{Spin}_{1,3}^{e}}^{\ell
}(M,\mathtt{g})$ and $\Phi\in\mathcal{C}\ell_{\mathrm{Spin}_{1,3}^{e}}%
^{r}(M,\mathtt{g})$, we denote the corresponding covariant derivatives by
$\nabla_{V}^{s}\Psi$ and $\nabla_{V}^{s}\Phi.$

We now recall some important formulas, without proof, concerning the covariant
derivatives of Clifford and spinor fields (for details see \cite{rodoliv2006}).

\begin{proposition}
The covariant derivative \emph{(}in a given gauge\emph{)} of a Clifford field
$A\in\mathcal{C}\ell(M,\mathtt{g}),$ in the direction of the vector field
$V\in\sec TM$ is given by
\begin{equation}
\nabla_{V}A=\partial_{V}\left(  A\right)  +\frac{1}{2}\left[  \omega
_{V},A\right]  , \label{s5}%
\end{equation}
where $\omega_{V}$ is the usual $\left(
{\textstyle\bigwedge\nolimits^{2}}
T^{\ast}M\text{-valued}\right)  $ connection $1$-form evaluated at the vector
field $V\in\sec TM$ written in the basis $\left\{  \theta_{\mathbf{a}%
}\right\}  $ and, if $A=A^{I}\theta_{I},$ then $\partial_{V}$ is the (Pfaff)
derivative operator such that $\partial_{V}\left(  A\right)  \equiv V\left(
A^{I}\right)  \theta_{I}.$
\end{proposition}

\begin{corollary}
The covariant derivative $\nabla_{V}$ on $\mathcal{C}\ell(M,\mathtt{g})$ acts
as a derivation on the algebra of sections, i.e., for $A,B\in\mathcal{C}%
\ell(M,\mathtt{g})$ and $V\in\sec TM,$ it holds%
\begin{equation}
\nabla_{V}\left(  AB\right)  =\nabla_{V}\left(  A\right)  B+A\nabla_{V}\left(
B\right)  \label{s6}%
\end{equation}

\end{corollary}

\begin{proposition}
Given $\Psi\in\mathcal{C}\ell_{\mathrm{Spin}_{1,3}^{e}}^{\ell}(M,\mathtt{g})$
and $\Phi\in\mathcal{C}\ell_{\mathrm{Spin}_{1,3}^{e}}^{r}(M,\mathtt{g})$ we
have,%
\begin{equation}
\nabla_{V}^{s}\Psi=\partial_{V}\left(  \Psi\right)  +\frac{1}{2}\omega_{V}%
\Psi, \label{s7}%
\end{equation}%
\begin{equation}
\nabla_{V}^{s}\Phi=\partial_{V}\left(  \Phi\right)  -\frac{1}{2}\Phi\omega
_{V}. \label{s8}%
\end{equation}

\end{proposition}

Now recalling that $\mathcal{C}\ell_{\mathrm{Spin}_{1,3}^{e}}^{\ell
}(M,\mathtt{g})$ ($\mathcal{C}\ell_{\mathrm{Spin}_{1,3}^{e}}^{r}%
(M,\mathtt{g})$) is a module over $\mathcal{C}\ell(M,\mathtt{g})$ \cite{LaMi},
we have the following proposition.

\begin{proposition}
Let $\nabla$ be the connection on $\mathcal{C}\ell(M,\mathtt{g})$ to which
$\nabla^{s}$ is related. Then, for any $V\in\sec TM$, $A\in\mathcal{C}%
\ell(M,\mathtt{g})$, $\Psi\in\mathcal{C}\ell_{\mathrm{Spin}_{1,3}^{e}}^{\ell
}(M,\mathtt{g})$ and $\Phi\in\mathcal{C}\ell_{\mathrm{Spin}_{1,3}^{e}}%
^{r}(M,\mathtt{g})$,%
\begin{equation}
\nabla_{V}^{s}\left(  A\Psi\right)  =A\nabla_{V}^{s}\left(  \Psi\right)
+\nabla_{V}\left(  A\right)  \Psi, \label{s9}%
\end{equation}%
\begin{equation}
\nabla_{V}^{s}\left(  A\Phi\right)  =\Phi\nabla_{V}\left(  A\right)
+\nabla_{V}^{s}\left(  \Phi\right)  A. \label{s10}%
\end{equation}

\end{proposition}

\begin{proposition}
\emph{(\cite{rodoliv2006})} Let $\mathbf{1}_{\Xi}^{r}\in\mathcal{C}%
\ell_{\mathrm{Spin}_{1,3}^{e}}^{r}(M,\mathtt{g})$ and $\mathbf{1}_{\Xi}^{\ell
}\in\mathcal{C}\ell_{\mathrm{Spin}_{1,3}^{e}}^{\ell}(M,\mathtt{g})$ be the
right and left unit section associated with spin coframe $\Xi.$ Then%
\begin{equation}
\nabla_{\mathbf{e}_{\mathbf{a}}}^{s}\mathbf{1}_{\Xi}^{r}=-\frac{1}%
{2}\mathbf{1}_{\Xi}^{r}\omega_{\mathbf{e}_{\mathbf{a}}},\qquad\nabla
_{\mathbf{e}_{\mathbf{a}}}^{s}\mathbf{1}_{\Xi}^{\ell}=\frac{1}{2}%
\omega_{\mathbf{e}_{\mathbf{a}}}\mathbf{1}_{\Xi}^{\ell}. \label{s11}%
\end{equation}

\end{proposition}

\subsection{Spin-Dirac Operator}

Let $\left\{  \theta^{\mathbf{a}}\right\}  \in\sec P_{\mathrm{SO}%
_{1,3}^{\mathbf{e}}}\left(  M\right)  ,$ such that for $\Xi\in\sec
P_{\mathrm{Spin}_{1,3}^{e}}\left(  M\right)  $, we have ( see, e.g.,
\cite{rodoliv2006}) $s:\sec P_{\mathrm{Spin}_{1,3}^{e}}\left(  M\right)
\rightarrow\sec P_{\mathrm{SO}_{1,3}^{\mathbf{e}}}\left(  M\right)  $ by
\[%
\begin{array}
[c]{c}%
s\left(  \Xi\right)  =\left\{  \theta^{\mathbf{a}}\right\}  ,\theta
^{\mathbf{a}}\in\mathcal{C}\ell(M,\mathtt{g}),\text{ }\theta^{\mathbf{a}%
}\left(  \mathbf{e}_{\mathbf{b}}\right)  =\delta_{\mathbf{b}}^{\mathbf{a}},\\
\theta^{\mathbf{a}}\theta^{\mathbf{b}}+\theta^{\mathbf{b}}\theta^{\mathbf{a}%
}=2\eta^{\mathbf{ab}},\qquad\mathbf{a,b}=0,1,2,3.
\end{array}
\]

\begin{definition}
The spin-Dirac operator acting on section of $\mathcal{C}\ell_{\mathrm{Spin}%
_{1,3}^{e}}^{\ell}(M,\mathtt{g})$ \emph{(}or $\mathcal{C}\ell_{\mathrm{Spin}%
_{1,3}^{e}}^{r}(M,\mathtt{g})$\emph{)} on a Riemann-Cartan spacetime is the
first order differential operator
\begin{equation}
\mbox{\boldmath$\partial$}^{s}=\theta^{\mathbf{\alpha}}\nabla_{\mathbf{e}%
_{\alpha}}^{s} \label{s12}%
\end{equation}
where $\{\mathbf{e}_{\alpha}\}$ and $\left\{  \theta^{\beta}\right\}  $ are
any pair of dual basis, and $\nabla_{\mathbf{e}_{\alpha}}^{s}$ is given by
\emph{Eqs.(\ref{s7}) }and \emph{(\ref{s8})}.
\end{definition}

\subsection{Representative of the spin-Dirac operator on $\mathcal{C}%
\ell(M,\mathtt{g})$}

In \cite{moro,rodoliv2006} it was shown in details that any spinor field
$\Psi\in\sec\mathcal{C}\ell_{\mathrm{Spin}_{1,3}^{e}}^{\ell}(M,\mathtt{g})$
can be represented (once a spin frame is selected) by a\footnote{$\mathcal{C}%
\ell^{(0)}(M,\mathtt{g})$ denotes the even subbundle of $\mathcal{C}%
\ell(M,\mathtt{g})$.} $\mathit{\psi}_{\Xi}\in\sec\mathcal{C}\ell
^{(0)}(M,\mathtt{g})$, called a representative of the spinor field in the
Clifford bundle, and such that
\begin{equation}
\mathit{\psi}_{\Xi}=\Psi1_{\Xi}^{r},
\end{equation}
with $1_{\Xi}^{r}\in\sec\mathcal{C}\ell_{\mathrm{Spin}_{1,3}^{e}}%
^{r}(M,\mathtt{g})$ a "unit" section. It was found that the representative of
$\mbox{\boldmath$\partial$}^{s}$ acting on $\Psi$ is
$\mbox{\boldmath$\partial$}^{(s)}=\theta^{\mathbf{a}}\nabla_{\mathbf{e}%
_{\mathbf{a}}}^{(s)}$ acting on $\mathit{\psi}_{\Xi}$ $\in\sec\mathcal{C}%
\ell(M,\mathtt{g})$ where $\mathbf{\nabla}_{V}^{(s)}$ is an \textquotedblleft
effective (spinorial) covariant\ derivative\textquotedblright\ acting on
$\mathit{\psi}_{\Xi}$ by%
\begin{equation}
\mathbf{\nabla}_{\mathbf{e}_{\mathbf{a}}}^{(s)}\mathit{\psi}_{\Xi
}:=\mathbf{\nabla}_{\mathbf{e}_{\mathbf{a}}}\mathit{\psi}_{\Xi}+{\frac{1}{2}%
}\mathit{\psi}_{\Xi}\omega_{\mathbf{e}_{\mathbf{a}}}, \label{hahaha}%
\end{equation}
from where it follows that
\begin{equation}
\mathbf{\nabla}_{\mathbf{e}_{\mathbf{a}}}^{(s)}\mathit{\psi}_{\Xi
}=\mathfrak{\partial}_{\mathbf{e}_{\mathbf{a}}}(\mathit{\psi}_{\Xi})+{\frac
{1}{2}}\omega_{\mathbf{e}_{\mathbf{a}}}\mathit{\psi}_{\Xi}\text{,}
\label{hahaha'}%
\end{equation}
which emulates the spinorial covariant derivative, as it should. We observe
moreover that if $\mathcal{C}\in\sec\mathcal{C}\ell(M,$\texttt{g}$)$ and if
$\mathit{\psi}_{\Xi}\in\sec\mathcal{C}\ell^{(0)}(M,$\texttt{g}$)$ is a
representative of a Dirac-Hestenes spinor field then%
\begin{equation}
\mathbf{\nabla}_{\mathtt{\ }\mathbf{e}_{\mathbf{a}}}^{(s)}\left(
\mathcal{C}\mathit{\psi}_{\Xi}\right)  =\left(  \mathbf{\nabla}_{\mathtt{\ }%
\mathbf{e}_{\mathbf{a}}}\mathcal{C}\right)  \mathit{\psi}_{\Xi}+\mathcal{C}%
\mathbf{\nabla}_{\mathtt{\ }\mathbf{e}_{\mathbf{a}}}^{(s)}\mathit{\psi}_{\Xi}.
\label{HAHAHA}%
\end{equation}

\section{Maxwell Equation on $\mathcal{C}\ell\left(  M,\mathtt{g}\right)  $
and on $\mathcal{C}\ell_{\mathrm{Spin}_{1,3}^{e}}^{r}(M,\mathtt{g})$}

As a useful example of the analogies and differences between the Clifford and
spin-Clifford bundles, we consider how to write the Maxwell \textit{equation}
in both formalisms.

The Maxwell equation in the Clifford bundle can be written, as well known
(see, e.g., \cite{rodoliv2006}),%
\begin{equation}
\mbox{\boldmath$\partial$} F=J_{\mathbf{e}} \label{m1}%
\end{equation}
where $F\in\sec%
{\displaystyle\bigwedge\nolimits^{2}}
T^{\ast}M\hookrightarrow\sec$\texttt{ }$\mathcal{C}\ell\left(  M,\mathtt{g}%
\right)  $ and $J_{\mathbf{e}}\in\sec%
{\displaystyle\bigwedge\nolimits^{1}}
T^{\ast}M\hookrightarrow\sec$\texttt{ }$\mathcal{C}\ell\left(  M,\mathtt{g}%
\right)  $.

Now, let $\psi_{\Xi}=F1_{\Xi}^{r}$ with $1_{\Xi}^{r}\in\sec\mathcal{C}%
\ell_{\mathrm{Spin}_{1,3}^{e}}^{r}(M,\mathtt{g})$. Then, recalling that
$\mathcal{C}\ell_{\mathrm{Spin}_{1,3}^{e}}^{r}(M,\mathtt{g})$ is a module over
$\mathcal{C}\ell(M,\mathtt{g}),$ we have $\psi_{\Xi}\in\sec\mathcal{C}%
\ell_{\mathrm{Spin}_{1,3}^{e}}^{r}(M,\mathtt{g})$, and recalling that
$\nabla_{\mathbf{e}_{\mathbf{a}}}^{s}1_{\Xi}^{r}=-\frac{1}{2}1_{\Xi}^{r}%
\omega_{\mathbf{e}_{\mathbf{a}}}$ we have%
\[
\nabla_{\mathbf{e}_{\mathbf{a}}}^{s}\psi_{\Xi}=\nabla_{\mathbf{e}_{\mathbf{a}%
}}^{s}\left(  F1_{\Xi}^{r}\right)  =\left(  \nabla_{\mathbf{e}_{\mathbf{a}}%
}F\right)  1_{\Xi}^{r}+F\left(  \nabla_{\mathbf{e}_{\mathbf{a}}}^{s}1_{\Xi
}^{r}\right)  =\left(  \nabla_{\mathbf{e}_{\mathbf{a}}}F\right)  1_{\Xi}%
^{r}-\frac{1}{2}F1_{\Xi}^{r}\omega_{\mathbf{e}_{\mathbf{a}}}%
\]
or%
\[
\theta^{\mathbf{a}}\nabla_{\mathbf{e}_{\mathbf{a}}}^{s}\psi_{\Xi}=\left(
\theta^{\mathbf{a}}\nabla_{\mathbf{e}_{\mathbf{a}}}F\right)  1_{\Xi}^{r}%
-\frac{1}{2}\theta^{\mathbf{a}}F1_{\Xi}^{r}\omega_{\mathbf{e}_{\mathbf{a}}}%
\]
from where
\[
\mbox{\boldmath$\partial$}^{s}\psi_{\Xi}=\left(  \mbox{\boldmath$\partial$}
F\right)  1_{\Xi}^{r}-\frac{1}{2}\theta^{\mathbf{a}}\psi_{\Xi}\omega
_{\mathbf{e}_{\mathbf{a}}}%
\]
and using Eq.(\ref{m1}) we end with%
\begin{equation}
\mbox{\boldmath$\partial$}^{s}\psi_{\Xi}+\frac{1}{2}\theta^{\mathbf{a}}%
\psi_{\Xi}\omega_{\mathbf{e}_{\mathbf{a}}}=J_{\mathbf{e}}1_{\Xi}^{r},
\label{m2}%
\end{equation}
where, of course, $J_{\mathbf{e}}1_{\Xi}^{r}\in\sec\mathcal{C}\ell
_{\mathrm{Spin}_{1,3}^{e}}^{r}(M,\mathtt{g})$. Eq.(\ref{m2}) is Maxwell
equation written in a spin-Clifford bundle, obviously equivalent to the
Maxwell equation written in the Clifford bundle.

Note that we can immediately recover Eq.(\ref{m1}) from Eq.(\ref{m2}). Indeed,
if $\psi_{\Xi}=F1_{\Xi}^{r}$ satisfies Eq.(\ref{m2}) we can write%
\[
\theta^{\mathbf{a}}\nabla_{\mathbf{e}_{\mathbf{a}}}^{s}\left(  F1_{\Xi}%
^{r}\right)  =J_{\mathbf{e}}1_{\Xi}^{r}-\frac{1}{2}\theta^{\mathbf{a}}F1_{\Xi
}^{r}\omega_{\mathbf{e}_{\mathbf{a}}}.
\]
Then%
\[
\left(  \theta^{\mathbf{a}}\nabla_{\mathbf{e}_{\mathbf{a}}}^{s}F\right)
1_{\Xi}^{r}+\theta^{\mathbf{a}}F\nabla_{\mathbf{e}_{\mathbf{a}}}^{s}1_{\Xi
}^{r}=J_{\mathbf{e}}1_{\Xi}^{r}+\theta^{\mathbf{a}}F\nabla_{\mathbf{e}%
_{\mathbf{a}}}^{s}1_{\Xi}^{r},
\]
from where%
\[
\left(  \mbox{\boldmath$\partial$} F\right)  1_{\Xi}^{r}=J_{\mathbf{e}}1_{\Xi
}^{r},
\]
and multiplying the above equation by $1_{\Xi}^{\ell}$ on the right, we obtain
the Eq. (\ref{m1}).

\section{The Square of the spin-Dirac Operator on a Riemann-Cartan Spacetime
and the Generalized Lichnerowicz Formula}

\subsection{Commutator of Covariant Derivatives of Spinor Fields}

In this section we the commutator of the \textit{representatives} of covariant
derivatives of spinor fields and the square of the spin-Dirac operator of a
Riemann-Cartan spacetime leading to the \ generalized Lichnerowicz formula.
Let $\psi\in\sec\mathcal{C\ell}^{(0)}(M,$\texttt{g}$)$ be a representative of
a \emph{DHSF }in a given spin frame $\Xi$ defining the orthonormal basis
$\{$\textbf{$e$}$_{\mathbf{a}}\}$ for\textbf{ }$TM$ and a corresponding dual
basis $\{\theta^{\mathbf{a}}\}$, $\theta^{\mathbf{a}}\in\sec%
{\displaystyle\bigwedge\nolimits^{1}}
T^{\ast}M\hookrightarrow\in\sec\mathcal{C\ell}(M,$\texttt{g}$)$. Let moreover,
$\{\theta_{\mathbf{a}}\}$ be the reciprocal basis of $\{\theta^{\mathbf{a}}%
\}$. We show that\footnote{Compare Eq.(\ref{comm deriv spin})\emph{ }with
Eq.(6.4.54) of Rammond's book \cite{ramond}, where there is a missing term.}
\begin{equation}
\lbrack\nabla_{\mathbf{e}_{\mathbf{a}}}^{(s)},\nabla_{\mathbf{e}_{b}}%
^{(s)}]\psi=\frac{1}{2}\mathfrak{R}(\theta_{\mathbf{a}}\wedge\theta
_{\mathbf{b}})\psi-(T_{\mathbf{ab}}^{\mathbf{c}}-\omega_{\mathbf{ab}%
}^{\mathbf{c}}+\omega_{\mathbf{ba}}^{\mathbf{c}})\nabla_{\mathbf{e}%
_{\mathbf{c}}}^{(s)}\psi, \label{comm deriv spin}%
\end{equation}
where the \textit{biform valued }curvature extensor field is (see details in
\cite{rodoliv2006}) is given by
\begin{equation}
\mathfrak{R}(u\wedge v)=\nabla_{\mathbf{u}}{\omega}_{\mathbf{v}}%
-\nabla_{\mathbf{v}}{\omega}_{\mathbf{u}}-\frac{1}{2}[{\omega}_{\mathbf{u}%
},{\omega}_{\mathbf{v}}]-{\omega}_{[\mathbf{u},\mathbf{v}]}, \label{curvature}%
\end{equation}
with $u=\mathbf{g}(\mathbf{u,}$ $)$, $v=\mathbf{g}(\mathbf{v,}$ $)$, and
$\mathbf{u,v\in}\sec TM$. Also ${\omega}_{\mathbf{v}}$ is the $\bigwedge
\nolimits^{2}T^{\ast}M$- valued connection (in the gauge defined by
$\{\theta_{\mathbf{a}}\}$) evaluated at $\mathbf{v}$.

Let us calculate $[\nabla_{\mathbf{u}}^{(s)},\nabla_{\mathbf{v}}^{(s)}]\psi$.
Taking into account that $\nabla_{\mathbf{u}}^{(s)}\psi=\nabla_{\mathbf{u}%
}\psi+\frac{1}{2}\psi\omega_{\mathbf{u}}$, we have%
\[
\nabla_{\mathbf{u}}^{(s)}\nabla_{\mathbf{v}}^{(s)}\psi=\nabla_{\mathbf{u}%
}\nabla_{\mathbf{v}}\psi+\frac{1}{2}(\nabla_{\mathbf{v}}\psi)\omega
_{\mathbf{u}}+\frac{1}{2}(\nabla_{\mathbf{u}}\psi)\omega_{\mathbf{v}}+\frac
{1}{4}\omega_{\mathbf{v}}\omega_{\mathbf{u}}+\frac{1}{2}\psi\nabla
_{\mathbf{u}}\omega_{\mathbf{v}}.
\]
Then,%
\begin{align}
\lbrack\nabla_{\mathbf{u}}^{(s)},\nabla_{\mathbf{v}}^{(s)}]\psi &
=[\nabla_{\mathbf{u}},\nabla_{\mathbf{v}}]\psi+\frac{1}{2}\psi(\nabla
_{\mathbf{u}}\omega_{\mathbf{v}}-\nabla_{\mathbf{v}}\omega_{\mathbf{u}}%
-\frac{1}{2}[\omega_{\mathbf{u}},\omega_{\mathbf{v}}])\nonumber\\
&  =\frac{1}{2}[\mathfrak{R}(u\wedge v),\psi]+\nabla_{\lbrack\mathbf{u,v}%
]}\psi+\frac{1}{2}\psi(\nabla_{\mathbf{u}}\omega_{\mathbf{v}}-\nabla
_{\mathbf{v}}\omega_{\mathbf{u}}-\frac{1}{2}[\omega_{\mathbf{u}}%
,\omega_{\mathbf{v}}])\nonumber\\
&  =\frac{1}{2}[\mathfrak{R}(u\wedge v),\psi]+\nabla_{\lbrack\mathbf{u,v}%
]}^{(s)}\psi-\frac{1}{2}\psi\omega_{\lbrack\mathbf{u,v]}}+\frac{1}{2}%
\psi(\nabla_{\mathbf{u}}\omega_{\mathbf{v}}-\nabla_{\mathbf{v}}\omega
_{\mathbf{u}}-\frac{1}{2}[\omega_{\mathbf{u}},\omega_{\mathbf{v}}])\nonumber\\
&  =\frac{1}{2}[\mathfrak{R}(u\wedge v),\psi]+\nabla_{\lbrack\mathbf{u,v}%
]}^{(s)}\psi+\frac{1}{2}\psi(\nabla_{\mathbf{u}}\omega_{\mathbf{v}}%
-\nabla_{\mathbf{v}}\omega_{\mathbf{u}}-\frac{1}{2}[\omega_{\mathbf{u}}%
,\omega_{\mathbf{v}}]-\omega_{\lbrack\mathbf{u,v]}})\nonumber\\
&  =\frac{1}{2}[\mathfrak{R}(u\wedge v),\psi]+\nabla_{\lbrack\mathbf{u,v}%
]}^{(s)}\psi+\frac{1}{2}\psi\mathfrak{R}(u\wedge v)\nonumber\\
&  =\frac{1}{2}\mathfrak{R}(u\wedge v)\psi+\nabla_{\lbrack\mathbf{u,v}]}%
^{(s)}\psi\text{.} \label{commut 2}%
\end{align}

From Eq.(\ref{commut 2}), the Eq.(\ref{comm deriv spin}) follows trivially.

\subsection{The Generalized Lichnerowicz Formula}

In this section we calculate the square of the spin-Dirac operator on a
Riemann-Cartan spacetime acting on a representative $\psi$ of the $\Psi\in
\sec\mathcal{C}\ell_{\mathrm{Spin}_{1,3}^{e}}^{\ell}(M,\mathtt{g})$.

\begin{proposition}%
\[
\left(  {\mbox{\boldmath$\partial$}}^{(s)}\right)  ^{2}\psi=\left(
\eta^{\mathbf{ab}}\nabla_{\mathbf{e}_{\mathbf{a}}}^{\left(  s\right)  }%
-\eta^{\mathbf{ac}}\omega_{\mathbf{ac}}^{\mathbf{b}}\right)  \nabla
_{\mathbf{e}_{\mathbf{b}}}^{\left(  s\right)  }\psi+\frac{1}{4}R\psi
+\mathbf{J}\psi-\Theta^{\mathbf{c}}\nabla_{\mathbf{e}_{\mathbf{c}}}^{(s)}\psi
\]

\end{proposition}

\begin{proof}
Taking notice that since $\theta^{\mathbf{b}}\in\sec\mathcal{C}\ell
(M,$\texttt{g}$)$, then $\nabla_{\mathbf{e}_{\mathbf{a}}}^{\left(  s\right)
}\theta^{\mathbf{b}}=\nabla_{\mathbf{e}_{\mathbf{a}}}\theta^{\mathbf{b}}$, we
have
\begin{equation}%
\begin{array}
[c]{ll}%
\left(  {\mbox{\boldmath$\partial$}}^{(s)}\right)  ^{2} & =\left(
\theta^{\mathbf{a}}\nabla_{\mathbf{e}_{\mathbf{a}}}^{\left(  s\right)
}\right)  \left(  \theta^{\mathbf{b}}\nabla_{\mathbf{e}_{\mathbf{b}}}^{\left(
s\right)  }\right) \\
& =\theta^{\mathbf{a}}\left[  \left(  \nabla_{\mathbf{e}_{\mathbf{a}}}%
\theta^{\mathbf{b}}\right)  \nabla_{\mathbf{e}_{\mathbf{b}}}^{\left(
s\right)  }+\theta^{\mathbf{b}}\nabla_{\mathbf{e}_{\mathbf{a}}}^{\left(
s\right)  }\nabla_{\mathbf{e}_{\mathbf{b}}}^{\left(  s\right)  }\right] \\
& =\theta^{\mathbf{a}}\mathbin\lrcorner\left[  \left(  \nabla_{\mathtt{\ }%
\mathbf{e}_{\mathbf{a}}}\theta^{\mathbf{b}}\right)  \nabla_{\mathbf{e}%
_{\mathbf{b}}}^{\left(  s\right)  }+\theta^{\mathbf{b}}\nabla_{\mathbf{e}%
_{\mathbf{a}}}^{\left(  s\right)  }\nabla_{\mathbf{e}_{\mathbf{b}}}^{\left(
s\right)  }\right] \\
& +\theta^{\mathbf{a}}\wedge\left[  \left(  \nabla_{\mathbf{e}_{\mathbf{a}}%
}\theta^{\mathbf{b}}\right)  \nabla_{\mathbf{e}_{\mathbf{b}}}^{\left(
s\right)  }+\theta^{\mathbf{b}}\nabla_{\mathbf{e}_{\mathbf{a}}}^{\left(
s\right)  }\nabla_{\mathbf{e}_{\mathbf{b}}}^{\left(  s\right)  }\right]
\end{array}
\label{b1}%
\end{equation}
and since $\nabla_{\mathbf{e}_{\mathbf{a}}}\theta^{\mathbf{b}}=-\omega
_{\mathbf{ac}}^{\mathbf{b}}\theta^{\mathbf{c}}$ we get after some algebra
\begin{equation}
\left(  {\mbox{\boldmath$\partial$}}^{(s)}\right)  ^{2}=\eta^{\mathbf{ab}%
}\left[  \nabla_{\mathbf{e}_{\mathbf{a}}}^{\left(  s\right)  }\nabla
_{\mathbf{e}_{\mathbf{b}}}^{\left(  s\right)  }-\omega_{\mathbf{ab}%
}^{\mathbf{c}}\nabla_{\mathbf{e}_{\mathbf{c}}}^{\left(  s\right)  }\right]
+\theta^{\mathbf{a}}\wedge\theta^{\mathbf{b}}\left[  \nabla_{\mathbf{e}%
_{\mathbf{a}}}^{\left(  s\right)  }\nabla_{\mathbf{e}_{\mathbf{b}}}^{\left(
s\right)  }-\omega_{\mathbf{ab}}^{\mathbf{c}}\nabla_{\mathbf{e}_{\mathbf{c}}%
}^{\left(  s\right)  }\right]  .
\end{equation}
Now, we define the operator%
\begin{equation}
{\mbox{\boldmath$\partial$}}^{(s)}\cdot{\mbox{\boldmath$\partial$}}^{(s)}%
=\eta^{\mathbf{ab}}\left[  \nabla_{\mathbf{e}_{\mathbf{a}}}^{\left(  s\right)
}\nabla_{\mathbf{e}_{\mathbf{b}}}^{\left(  s\right)  }-\omega_{\mathbf{ab}%
}^{\mathbf{c}}\nabla_{\mathbf{e}_{\mathbf{c}}}^{\left(  s\right)  }\right]  ,
\label{generalized laplacian}%
\end{equation}
which may be called the \textit{generalized spin Dalembertian,} and the
operator
\begin{equation}
{\mbox{\boldmath$\partial$}}^{(s)}\wedge{\mbox{\boldmath$\partial$}}%
^{(s)}=\theta^{\mathbf{a}}\wedge\theta^{\mathbf{b}}\left[  \nabla
_{\mathbf{e}_{\mathbf{a}}}^{\left(  s\right)  }\nabla_{\mathbf{e}_{\mathbf{b}%
}}^{\left(  s\right)  }-\omega_{\mathbf{ab}}^{\mathbf{c}}\nabla_{\mathbf{e}%
_{\mathbf{c}}}^{\left(  s\right)  }\right]  , \label{generalized ricci}%
\end{equation}
which will be called \textit{twisted curvature operator}. Then, we can write%
\begin{equation}
\left(  {\mbox{\boldmath$\partial$}}^{(s)}\right)  ^{2}%
={\mbox{\boldmath$\partial$}}^{(s)}\cdot{\mbox{\boldmath$\partial$}}%
^{(s)}+{\mbox{\boldmath$\partial$}}^{(s)}\wedge{\mbox{\boldmath$\partial$}}%
^{(s)}. \label{decomposition}%
\end{equation}
On the other hand, we have%
\begin{equation}
{\mbox{\boldmath$\partial$}}^{(s)}\cdot{\mbox{\boldmath$\partial$}}%
^{(s)}=\left[  \eta^{\mathbf{ab}}\nabla_{\mathbf{e}_{\mathbf{a}}}^{\left(
s\right)  }-\eta^{\mathbf{ac}}\omega_{\mathbf{ac}}^{\mathbf{b}}\right]
\nabla_{\mathbf{e}_{\mathbf{b}}}^{\left(  s\right)  }, \label{cc1}%
\end{equation}
and%
\begin{equation}%
\begin{array}
[c]{l}%
{\mbox{\boldmath$\partial$}}^{(s)}\wedge{\mbox{\boldmath$\partial$}}%
^{(s)}=\frac{1}{2}{\mbox{\boldmath$\partial$}}^{(s)}\wedge
{\mbox{\boldmath$\partial$}}^{(s)}+\frac{1}{2}{\mbox{\boldmath$\partial$}}%
^{(s)}\wedge{\mbox{\boldmath$\partial$}}^{(s)}\\
=\frac{1}{2}\theta^{\mathbf{a}}\wedge\theta^{\mathbf{b}}\left[  \nabla
_{\mathbf{e}_{\mathbf{a}}}^{\left(  s\right)  }\nabla_{\mathbf{e}_{\mathbf{b}%
}}^{\left(  s\right)  }-\omega_{\mathbf{ab}}^{\mathbf{c}}\nabla_{\mathbf{e}%
_{\mathbf{c}}}^{\left(  s\right)  }\right]  +\frac{1}{2}\theta^{\mathbf{b}%
}\wedge\theta^{\mathbf{a}}\left[  \nabla_{\mathbf{e}_{\mathbf{b}}}^{\left(
s\right)  }\nabla_{\mathbf{e}_{\mathbf{a}}}^{\left(  s\right)  }%
-\omega_{\mathbf{ba}}^{\mathbf{c}}\nabla_{\mathbf{e}_{\mathbf{c}}}^{\left(
s\right)  }\right] \\
=\frac{1}{2}\theta^{\mathbf{a}}\wedge\theta^{\mathbf{b}}\left[  \nabla
_{\mathbf{e}_{\mathbf{a}}}^{\left(  s\right)  }\nabla_{\mathbf{e}_{\mathbf{b}%
}}^{\left(  s\right)  }-\nabla_{\mathbf{e}_{\mathbf{b}}}^{\left(  s\right)
}\nabla_{\mathbf{e}_{\mathbf{a}}}^{\left(  s\right)  }-\left(  \omega
_{\mathbf{ab}}^{\mathbf{c}}-\omega_{\mathbf{ba}}^{\mathbf{c}}\right)
\nabla_{\mathbf{e}_{\mathbf{c}}}^{\left(  s\right)  }\right] \\
=\frac{1}{2}\theta^{\mathbf{a}}\wedge\theta^{\mathbf{b}}\left[  \nabla
_{\mathbf{e}_{\mathbf{a}}}^{\left(  s\right)  }\nabla_{\mathbf{e}_{\mathbf{b}%
}}^{\left(  s\right)  }-\nabla_{\mathbf{e}_{\mathbf{b}}}^{\left(  s\right)
}\nabla_{\mathbf{e}_{\mathbf{a}}}^{\left(  s\right)  }-(c_{\mathbf{ab}%
}^{\mathbf{c}}+T_{\mathbf{ab}}^{\mathbf{c}})\nabla_{\mathbf{e}_{\mathbf{c}}%
}^{\left(  s\right)  }\right]  .
\end{array}
\label{cc2}%
\end{equation}
Taking into account that $T_{\mathbf{ab}}^{\mathbf{c}}=\omega_{\mathbf{ab}%
}^{\mathbf{c}}-\omega_{\mathbf{ba}}^{\mathbf{c}}-c_{\mathbf{ab}}^{\mathbf{c}}%
$, we have from Eq.(\ref{cc1}) and Eq.(\ref{cc2}) that
\begin{equation}
\left(  {\mbox{\boldmath$\partial$}}^{(s)}\right)  ^{2}=\left[  \eta
^{\mathbf{ab}}\nabla_{\mathbf{e}_{\mathbf{a}}}^{\left(  s\right)  }%
-\eta^{\mathbf{ac}}\omega_{\mathbf{ac}}^{\mathbf{b}}\right]  \nabla
_{\mathbf{e}_{\mathbf{b}}}^{\left(  s\right)  }+\frac{1}{2}\theta^{\mathbf{a}%
}\wedge\theta^{\mathbf{b}}\left[  \nabla_{\mathbf{e}_{\mathbf{a}}}^{\left(
s\right)  }\nabla_{\mathbf{e}_{\mathbf{b}}}^{\left(  s\right)  }%
-\nabla_{\mathbf{e}_{\mathbf{b}}}^{\left(  s\right)  }\nabla_{\mathbf{e}%
_{\mathbf{a}}}^{\left(  s\right)  }-(c_{\mathbf{ab}}^{\mathbf{c}%
}+T_{\mathbf{ab}}^{\mathbf{c}})\nabla_{\mathbf{e}_{\mathbf{c}}}^{\left(
s\right)  }\right]  . \label{cc3}%
\end{equation}
On the other hand, from Eq.(\ref{comm deriv spin}), we have%
\[
\left[  \nabla_{\mathbf{e}_{\mathbf{a}}}^{\left(  s\right)  },\nabla
_{\mathbf{e}_{\mathbf{b}}}^{\left(  s\right)  }\right]  \psi=\frac{1}%
{2}\mathcal{R}\left(  \theta_{\mathbf{a}}\wedge\theta_{\mathbf{b}}\right)
\psi+c_{\mathbf{ab}}^{\mathbf{c}}\nabla_{\mathtt{\ }\mathbf{e}_{\mathbf{c}}%
}^{(s)}\psi
\]
and then Eq.(\ref{cc3}) becomes
\begin{align*}
\left(  {\mbox{\boldmath$\partial$}}^{(s)}\right)  ^{2}\psi &  =\left[
\eta^{\mathbf{ab}}\nabla_{\mathbf{e}_{\mathbf{a}}}^{\left(  s\right)  }%
-\eta^{\mathbf{ac}}\omega_{\mathbf{ac}}^{\mathbf{b}}\right]  \nabla
_{\mathtt{\ }\mathbf{e}_{\mathbf{b}}}^{\left(  s\right)  }\psi+\frac{1}%
{4}(\theta^{\mathbf{a}}\wedge\theta^{\mathbf{b}})\mathcal{R}\left(
\theta_{\mathbf{a}}\wedge\theta_{\mathbf{b}}\right)  \psi\\
&  -\frac{1}{2}\theta^{\mathbf{a}}\wedge\theta^{\mathbf{b}}T_{\mathbf{ab}%
}^{\mathbf{c}}\nabla_{\mathbf{e}_{\mathbf{c}}}^{(s)}\psi\\
&  =\left[  \eta^{\mathbf{ab}}\nabla_{\mathbf{e}_{\mathbf{a}}}^{\left(
s\right)  }-\eta^{\mathbf{ac}}\omega_{\mathbf{ac}}^{\mathbf{b}}\right]
\nabla_{\mathbf{e}_{\mathbf{b}}}^{\left(  s\right)  }\psi+\frac{1}{4}%
(\theta^{\mathbf{a}}\wedge\theta^{\mathbf{b}})\mathcal{R}\left(
\theta_{\mathbf{a}}\wedge\theta_{\mathbf{b}}\right)  \psi-\Theta^{\mathbf{c}%
}\nabla_{\mathbf{e}_{\mathbf{c}}}^{(s)}\psi.
\end{align*}

We need to compute $(\theta^{\mathbf{a}}\wedge\theta^{\mathbf{b}}%
)\mathcal{R}\left(  \theta_{\mathbf{a}}\wedge\theta_{\mathbf{b}}\right)  $. We
have
\begin{align*}
(\theta^{\mathbf{a}}\wedge\theta^{\mathbf{b}})\mathcal{R}\left(
\theta_{\mathbf{a}}\wedge\theta_{\mathbf{b}}\right)   &  =\langle
(\theta^{\mathbf{a}}\wedge\theta^{\mathbf{b}})\mathcal{R}\left(
\theta_{\mathbf{a}}\wedge\theta_{\mathbf{b}}\right)  \rangle_{0}%
+\langle(\theta^{\mathbf{a}}\wedge\theta^{\mathbf{b}})\mathcal{R}\left(
\theta_{\mathbf{a}}\wedge\theta_{\mathbf{b}}\right)  \rangle_{2}\\
&  +\langle(\theta^{\mathbf{a}}\wedge\theta^{\mathbf{b}})\mathcal{R}\left(
\theta_{\mathbf{a}}\wedge\theta_{\mathbf{b}}\right)  \rangle_{4}.
\end{align*}

Now, \ we get
\begin{align*}
\langle(\theta^{\mathbf{a}}\wedge\theta^{\mathbf{b}})\mathcal{R}\left(
\theta_{\mathbf{a}}\wedge\theta_{\mathbf{b}}\right)  \rangle_{0}  &
:=(\theta^{\mathbf{a}}\wedge\theta^{\mathbf{b}})\mathbin\lrcorner
\mathcal{R}\left(  \theta_{\mathbf{a}}\wedge\theta_{\mathbf{b}}\right) \\
&  =-(\theta^{\mathbf{a}}\wedge\theta^{\mathbf{b}})\cdot\mathcal{R}\left(
\theta_{\mathbf{a}}\wedge\theta_{\mathbf{b}}\right)  =R.
\end{align*}

Also,
\[
\langle(\theta^{\mathbf{a}}\wedge\theta^{\mathbf{b}})\mathcal{R}\left(
\theta_{\mathbf{a}}\wedge\theta_{\mathbf{b}}\right)  \rangle_{2}%
=\theta^{\mathbf{a}}\wedge(\theta^{\mathbf{b}}\mathbin\lrcorner\mathcal{R}%
\left(  \theta_{\mathbf{a}}\wedge\theta_{\mathbf{b}}\right)  )+\theta
^{\mathbf{a}}\mathbin\lrcorner(\theta^{\mathbf{b}}\wedge\mathcal{R}\left(
\theta_{\mathbf{a}}\wedge\theta_{\mathbf{b}}\right)  )
\]
and recalling the identity (see \cite{rodoliv2006})
\[
\theta^{\mathbf{a}}\mathbin\lrcorner(\theta^{\mathbf{b}}\wedge\mathcal{R}%
\left(  \theta_{\mathbf{a}}\wedge\theta_{\mathbf{b}}\right)  )-\theta
^{\mathbf{a}}\wedge(\theta^{\mathbf{b}}\mathbin\lrcorner\mathcal{R}\left(
\theta_{\mathbf{a}}\wedge\theta_{\mathbf{b}}\right)  )=(\theta^{\mathbf{a}%
}\cdot\theta^{\mathbf{b}})\mathcal{R}\left(  \theta_{\mathbf{a}}\wedge
\theta_{\mathbf{b}}\right)  ,
\]
it follows that
\[
\langle(\theta^{\mathbf{a}}\wedge\theta^{\mathbf{b}})\mathcal{R}\left(
\theta_{\mathbf{a}}\wedge\theta_{\mathbf{b}}\right)  \rangle_{2}%
=(\theta^{\mathbf{a}}\cdot\theta^{\mathbf{b}})\mathcal{R}\left(
\theta_{\mathbf{a}}\wedge\theta_{\mathbf{b}}\right)  =\eta^{\mathbf{ab}%
}\mathcal{R}\left(  \theta_{\mathbf{a}}\wedge\theta_{\mathbf{b}}\right)  =0.
\]

It remains to calculate $\langle\theta^{\mathbf{a}}\wedge\theta^{\mathbf{b}%
}\mathcal{R}\left(  \theta_{\mathbf{a}}\wedge\theta_{\mathbf{b}}\right)
\rangle_{4}$. We have%
\begin{align*}
\langle\theta^{\mathbf{a}}\wedge\theta^{\mathbf{b}}\mathcal{R}\left(
\theta_{\mathbf{a}}\wedge\theta_{\mathbf{b}}\right)  \rangle_{4}  &
=\theta^{\mathbf{a}}\wedge\theta^{\mathbf{b}}\wedge\mathcal{R}\left(
\theta_{\mathbf{a}}\wedge\theta_{\mathbf{b}}\right)  =\frac{1}{2}%
R_{\mathbf{abcd}}\theta^{\mathbf{a}}\wedge\theta^{\mathbf{b}}\wedge
\theta^{\mathbf{c}}\wedge\theta^{\mathbf{d}}\\
&  =\frac{1}{6}(R_{\mathbf{abcd}}\mathbf{\theta}^{\mathbf{abcd}}%
+R_{\mathbf{acdb}}\mathbf{\theta}^{\mathbf{acdb}}+R_{\mathbf{adbc}%
}\mathbf{\theta}^{\mathbf{adbc}})\\
&  =\frac{1}{6}(R_{\mathbf{abcd}}+R_{\mathbf{acdb}}+R_{\mathbf{adbc}}%
)\theta^{\mathbf{abcd}}.
\end{align*}
Now, we recall a well known result (see, e.g., \cite{rodoliv2006} )
\[
R_{\mathbf{abcd}}=\mathring{R}_{\mathbf{abcd}}+J_{\mathbf{ab[cd]}}%
\]
where $\mathring{R}_{\mathbf{abcd}}$ are the components of the Riemann tensor
of the Levi-Civita connection of $\mathbf{g}$ and
\begin{align*}
J_{\mathbf{a\,cd}}^{\;\mathbf{b}}  &  =\nabla_{\mathbf{c}}K_{\mathbf{da}%
}^{\mathbf{b}}-K_{\mathbf{ck}}^{\mathbf{b}}K_{\mathbf{da}}^{\mathbf{k}%
}+K_{\mathbf{cd}}^{\mathbf{k}}K_{\mathbf{ka}}^{\mathbf{b}},\\
J_{\mathbf{a\,[cd]}}^{\;\mathbf{b}}  &  =J_{\mathbf{a\,cd}}^{\;\mathbf{b}%
}-J_{\mathbf{a\,dc}}^{\;\mathbf{b}},
\end{align*}
with $K_{\mathbf{cd}}^{\mathbf{k}}$ given by
\[
K_{\mathbf{cd}}^{\mathbf{k}}=-\frac{1}{2}\eta^{\mathbf{km}}(\eta_{\mathbf{nc}%
}T_{\mathbf{md}}^{\mathbf{n}}+\eta_{\mathbf{nd}}T_{\mathbf{mc}}^{\mathbf{n}%
}-\eta_{\mathbf{nm}}T_{\mathbf{cd}}^{\mathbf{n}}).
\]

Moreover, taking into account the (well known) first Bianchi identity,
$\mathring{R}_{\mathbf{abcd}}+\mathring{R}_{\mathbf{acdb}}+\mathring
{R}_{\mathbf{adbc}}=0$, we have%

\begin{equation}
\left(  {\mbox{\boldmath$\partial$}}^{(s)}\right)  ^{2}\psi=\left[
\eta^{\mathbf{ab}}\nabla_{\mathbf{e}_{\mathbf{a}}}^{\left(  s\right)  }%
-\eta^{\mathbf{ac}}\omega_{\mathbf{ac}}^{\mathbf{b}}\right]  \nabla
_{\mathbf{e}_{\mathbf{b}}}^{\left(  s\right)  }\psi+\frac{1}{4}R\psi
+\mathbf{J}\psi-\Theta^{\mathbf{c}}\nabla_{\mathbf{e}_{\mathbf{c}}}^{(s)}\psi,
\label{square spin dirac}%
\end{equation}

where%
\begin{align}
\mathbf{J}  &  =\frac{1}{6}(J_{\mathbf{ab[cd]}}+J_{\mathbf{ac[db]}%
}+J_{\mathbf{ad[bc]}})\theta^{\mathbf{abcd}}\nonumber\\
&  =\frac{1}{6}(J_{\mathbf{ab[cd]}}+J_{\mathbf{ac[db]}}+J_{\mathbf{ad[bc]}%
})\epsilon_{\mathbf{0123}}^{\mathbf{abcd}}\tau_{\mathtt{\mathbf{g}}},
\label{J}%
\end{align}
and the proposition is proved.
\end{proof}

\begin{remark}
\emph{Eq.(\ref{square spin dirac}) }may be called the generalized Lichnerowicz
formula and \emph{(}equivalent expressions\emph{)} appears in the case of a
totally skew-symmetric torsion in many different contexts, like, e.g., in the
geometry of moduli spaces of a class of black holes, the geometry of NS-5
brane solutions of type II supergravity theories and BPS solitons in some
string theories \emph{(\cite{dalakov})} and many \ important topics of modern
mathematics \emph{(}see \emph{\cite{bismut,friedrich})}. For a \ Levi-Civita
connection we have that $\mathbf{J}=0$ and $\Theta^{\mathbf{c}}=0$ and we
obtain the famous Lichnerowicz formula \emph{\cite{lich}}.
\end{remark}

\section{Relation Between the Square of the Spin-Dirac Operator and the Dirac
Operator}

In this section taking advantage that any $\psi\in\sec\mathcal{C}%
\ell_{\mathrm{Spin}_{1,3}^{e}}^{\ell}(M,\mathtt{g})$ can be $\psi
=A1_{\mathbf{\Xi}}^{\ell}$ with $A\in\sec\mathcal{C}\ell(M,\mathtt{g})$ and
$1_{\mathbf{\Xi}}^{\ell}\in\sec\mathcal{C}\ell_{\mathrm{Spin}_{1,3}^{e}}%
^{\ell}(M,\mathtt{g})$ we find two noticeable formulas: the first relates the
square of the \textit{spin-Dirac operator }($\theta^{\mathbf{a}}%
\nabla_{\mathbf{e}_{\mathbf{a}}}^{s}$) acting on $\psi$ with the square of the
Dirac operator ($\theta^{\mathbf{a}}\nabla_{\mathbf{e}_{\mathbf{a}}}$) acting
on $A$ associated with the covariant derivative $\nabla$ of a Riemann-Cartan
spacetime $(M,\mathbf{g},\nabla)$ admitting a spin structure; the second
formula relates the square of the \textit{spin-Dirac operator }($\theta
^{\mathbf{a}}\nabla_{\mathbf{e}_{\mathbf{a}}}^{s}$) acting on $\psi$ with the
square of the standard Dirac operator ($\theta^{\mathbf{a}}D_{\mathbf{e}%
_{\mathbf{a}}}$) associated with the covariant derivative $D$ of a Lorentzian
spacetime $(M,\mathbf{g},D)$.

We already know that if $\psi\in\sec\mathcal{C}\ell_{\mathrm{Spin}_{1,3}^{e}%
}^{\ell}(M,\mathtt{g})$, then%
\[
\nabla_{\mathbf{e}_{\mathbf{a}}}^{s}\psi=\partial_{\mathbf{e}_{\mathbf{a}}%
}\psi+\frac{1}{2}\omega_{\mathbf{e}_{\mathbf{a}}}\psi
\]
and that if $\phi\in\sec\mathcal{C}\ell_{\mathrm{Spin}_{1,3}^{e}}%
^{r}(M,\mathtt{g})$ then
\[
\nabla_{\mathbf{e}_{\mathbf{a}}}^{s}\phi=\partial_{\mathbf{e}_{\mathbf{a}}%
}\phi-\frac{1}{2}\phi\omega_{\mathbf{e}_{\mathbf{a}}}.
\]

On the other hand, we recall that a direct calculation gives%
\begin{equation}
\left(  \mbox{\boldmath$\partial$}^{s}\right)  ^{2}=\eta^{\mathbf{ab}}\left[
\nabla_{\mathbf{e}_{\mathbf{a}}}^{s}\nabla_{\mathbf{e}_{\mathbf{b}}}%
^{s}-\omega_{\mathbf{ab}}^{\mathbf{c}}\nabla_{\mathbf{e}_{\mathbf{c}}}%
^{s}\right]  +\theta^{\mathbf{a}}\wedge\theta^{\mathbf{b}}\left[
\nabla_{\mathbf{e}_{\mathbf{a}}}^{s}\nabla_{\mathbf{e}_{\mathbf{b}}}%
^{s}-\omega_{\mathbf{ab}}^{\mathbf{c}}\nabla_{\mathbf{e}_{\mathbf{c}}}%
^{s}\right]  . \label{r8}%
\end{equation}
Now, to calculate $\nabla_{\mathbf{e}_{\mathbf{a}}}^{s}$\ in term of
$\nabla_{\mathbf{e}_{\mathbf{a}}}$, we must first recall that the domains of
these operators are different. Let $A\in\sec\mathcal{C}\ell(M,\mathtt{g})$ and
let $1_{\mathbf{\Xi}}^{\ell}\in\sec\mathcal{C}\ell_{\mathrm{Spin}_{1,3}^{e}%
}^{\ell}(M,\mathtt{g})$. Then $\psi=A1_{\mathbf{\Xi}}^{\ell}\in\sec
\mathcal{C}\ell_{\mathrm{Spin}_{1,3}^{e}}^{\ell}(M,\mathtt{g})$ and we can
write%
\[%
\begin{array}
[c]{ll}%
\nabla_{\mathbf{e}_{\mathbf{c}}}^{s}\psi & =\nabla_{\mathbf{e}_{\mathbf{c}}%
}^{s}\left(  A1_{\mathbf{\Xi}}^{\ell}\right)  =\partial_{\mathbf{e}%
_{\mathbf{c}}}\left(  A1_{\mathbf{\Xi}}^{\ell}\right)  +\frac{1}{2}%
\omega_{\mathbf{e}_{\mathbf{c}}}A1_{\mathbf{\Xi}}^{\ell}\medskip\\
& =\left(  \partial_{\mathbf{e}_{\mathbf{c}}}A\right)  1_{\mathbf{\Xi}}^{\ell
}+A\partial_{\mathbf{e}_{\mathbf{c}}}1_{\mathbf{\Xi}}^{\ell}+\frac{1}{2}%
\omega_{\mathbf{e}_{\mathbf{c}}}A1_{\mathbf{\Xi}}^{\ell}-\frac{1}{2}%
A\omega_{\mathbf{e}_{\mathbf{c}}}1_{\mathbf{\Xi}}^{\ell}+\frac{1}{2}%
A\omega_{\mathbf{e}_{\mathbf{c}}}1_{\mathbf{\Xi}}^{\ell}\medskip\\
& =\left(  \partial_{\mathbf{e}_{\mathbf{c}}}A\right)  1_{\mathbf{\Xi}}^{\ell
}+\frac{1}{2}\omega_{\mathbf{e}_{\mathbf{c}}}A1_{\mathbf{\Xi}}^{\ell}-\frac
{1}{2}A\omega_{\mathbf{e}_{\mathbf{c}}}1_{\mathbf{\Xi}}^{\ell}+\frac{1}%
{2}A\omega_{\mathbf{e}_{\mathbf{c}}}1_{\mathbf{\Xi}}^{\ell}\medskip\\
& =\left(  \partial_{\mathbf{e}_{\mathbf{c}}}A+\frac{1}{2}\omega
_{\mathbf{e}_{\mathbf{c}}}A-\frac{1}{2}A\omega_{\mathbf{e}_{\mathbf{c}}%
}\right)  1_{\mathbf{\Xi}}^{\ell}+\frac{1}{2}A\omega_{\mathbf{e}_{\mathbf{c}}%
}1_{\mathbf{\Xi}}^{\ell}.
\end{array}
\]
Then,%

\begin{equation}
\nabla_{\mathbf{e}_{\mathbf{c}}}^{s}\psi=\left(  \nabla_{\mathbf{e}%
_{\mathbf{c}}}A\right)  1_{\mathbf{\Xi}}^{\ell}+\frac{1}{2}A\omega
_{\mathbf{e}_{\mathbf{c}}}1_{\mathbf{\Xi}}^{\ell}, \label{r9}%
\end{equation}
where we notice that $\partial_{\mathbf{e}_{\mathbf{c}}}1_{\mathbf{\Xi}}%
^{\ell}=0$. Using Eq.(\ref{r9})\ we have%
\begin{equation}%
\begin{array}
[c]{ll}%
\nabla_{\mathbf{e}_{\mathbf{a}}}^{s}\nabla_{\mathbf{e}_{\mathbf{b}}}^{s}\psi &
=\nabla_{\mathbf{e}_{\mathbf{a}}}^{s}\left(  \left(  \nabla_{\mathbf{e}%
_{\mathbf{b}}}A\right)  1_{\mathbf{\Xi}}^{\ell}+\frac{1}{2}A\omega
_{\mathbf{e}_{\mathbf{b}}}1_{\mathbf{\Xi}}^{\ell}\right)  \medskip\\
& =\left(  \nabla_{\mathbf{e}_{\mathbf{a}}}\nabla_{\mathbf{e}_{\mathbf{b}}%
}A\right)  1_{\mathbf{\Xi}}^{\ell}+\left(  \nabla_{\mathbf{e}_{\mathbf{b}}%
}A\right)  \nabla_{\mathbf{e}_{\mathbf{a}}}^{s}1_{\mathbf{\Xi}}^{\ell}%
+\frac{1}{2}\nabla_{\mathbf{e}_{\mathbf{a}}}^{s}\left(  A\omega_{\mathbf{e}%
_{\mathbf{b}}}1_{\mathbf{\Xi}}^{\ell}\right)  \medskip\\
& =\left(  \nabla_{\mathbf{e}_{\mathbf{a}}}\nabla_{\mathbf{e}_{\mathbf{b}}%
}A\right)  1_{\mathbf{\Xi}}^{\ell}+\frac{1}{2}\left(  \nabla_{\mathbf{e}%
_{\mathbf{b}}}A\right)  \omega_{\mathbf{e}_{\mathbf{a}}}1_{\mathbf{\Xi}}%
^{\ell}+\frac{1}{2}\nabla_{\mathbf{e}_{\mathbf{a}}}^{s}\left(  A\omega
_{\mathbf{e}_{\mathbf{b}}}1_{\mathbf{\Xi}}^{\ell}\right)  ,
\end{array}
\label{r10}%
\end{equation}
and using Eq.(\ref{r9}) and Eq.(\ref{r10}) we have%
\begin{equation}%
\begin{array}
[c]{ll}%
\left(  \nabla_{\mathbf{e}_{\mathbf{a}}}^{s}\nabla_{\mathbf{e}_{\mathbf{b}}%
}^{s}-\omega_{\mathbf{ab}}^{\mathbf{c}}\nabla_{\mathbf{e}_{\mathbf{c}}}%
^{s}\right)  \psi & =\left(  \nabla_{\mathbf{e}_{\mathbf{a}}}\nabla
_{\mathbf{e}_{\mathbf{b}}}A-\omega_{\mathbf{ab}}^{\mathbf{c}}\nabla
_{\mathbf{e}_{\mathbf{c}}}A\right)  1_{\mathbf{e}}^{\ell}+\frac{1}{2}\left(
\nabla_{\mathbf{e}_{\mathbf{b}}}A\right)  \omega_{\mathbf{e}_{\mathbf{a}}%
}1_{\mathbf{e}}^{\ell}\medskip\\
& +\frac{1}{2}\nabla_{\mathbf{e}_{\mathbf{a}}}^{s}\left(  A\omega
_{\mathbf{e}_{\mathbf{b}}}1_{\mathbf{e}}^{\ell}\right)  -\frac{1}{2}%
\omega_{\mathbf{ab}}^{\mathbf{c}}A\omega_{\mathbf{e}_{\mathbf{c}}%
}1_{\mathbf{e}}^{\ell}.
\end{array}
\label{r11}%
\end{equation}
Substituting Eq.(\ref{r11}) into Eq.(\ref{r8}) we obtain%
\[%
\begin{array}
[c]{ll}%
\left(  \mbox{\boldmath$\partial$}^{s}\right)  ^{2}\psi & =\eta^{\mathbf{ab}%
}\left[  \left(  \nabla_{\mathbf{e}_{\mathbf{a}}}\nabla_{\mathbf{e}%
_{\mathbf{b}}}A-\omega_{\mathbf{ab}}^{\mathbf{c}}\nabla_{\mathbf{e}%
_{\mathbf{c}}}A\right)  1_{\mathbf{\Xi}}^{\ell}+\frac{1}{2}\left(
\nabla_{\mathbf{e}_{\mathbf{b}}}A\right)  \omega_{\mathbf{e}_{\mathbf{a}}%
}1_{\mathbf{\Xi}}^{\ell}\right.  \medskip\\
& \left.  +\frac{1}{2}\nabla_{\mathbf{e}_{\mathbf{a}}}^{s}\left(
A\omega_{\mathbf{e}_{\mathbf{b}}}1_{\mathbf{\Xi}}^{\ell}\right)  -\frac{1}%
{2}\omega_{\mathbf{ab}}^{\mathbf{c}}A\omega_{\mathbf{e}_{\mathbf{c}}%
}1_{\mathbf{\Xi}}^{\ell}\right]  \medskip\\
& +\theta^{\mathbf{a}}\wedge\theta^{\mathbf{b}}\left[  \left(  \nabla
_{\mathbf{e}_{\mathbf{a}}}\nabla_{\mathbf{e}_{\mathbf{b}}}A-\omega
_{\mathbf{ab}}^{\mathbf{c}}\nabla_{\mathbf{e}_{\mathbf{c}}}A\right)
1_{\mathbf{\Xi}}^{\ell}+\frac{1}{2}\left(  \nabla_{\mathbf{e}_{\mathbf{b}}%
}A\right)  \omega_{\mathbf{e}_{\mathbf{a}}}1_{\mathbf{\Xi}}^{\ell}\right.
\medskip\\
& \left.  +\frac{1}{2}\nabla_{\mathbf{e}_{\mathbf{a}}}^{s}\left(
A\omega_{\mathbf{e}_{\mathbf{b}}}1_{\mathbf{\Xi}}^{\ell}\right)  -\frac{1}%
{2}\omega_{\mathbf{ab}}^{\mathbf{c}}A\omega_{\mathbf{e}_{\mathbf{c}}%
}1_{\mathbf{\Xi}}^{\ell}\right]  ,
\end{array}
\]
or%
\begin{equation}%
\begin{array}
[c]{ll}%
\left(  \mbox{\boldmath$\partial$}^{s}\right)  ^{2}\psi & =\left(
\mbox{\boldmath$\partial$}^{2}A\right)  1_{\mathbf{\Xi}}^{\ell}+\eta
^{\mathbf{ab}}\left[  \frac{1}{2}\left(  \nabla_{\mathbf{e}_{\mathbf{b}}%
}A\right)  \omega_{\mathbf{e}_{\mathbf{a}}}1_{\mathbf{\Xi}}^{\ell}+\frac{1}%
{2}\nabla_{\mathbf{e}_{\mathbf{a}}}^{s}\left(  A\omega_{\mathbf{e}%
_{\mathbf{b}}}1_{\mathbf{\Xi}}^{\ell}\right)  -\frac{1}{2}\omega_{\mathbf{ab}%
}^{\mathbf{c}}A\omega_{\mathbf{e}_{\mathbf{c}}}1_{\mathbf{\Xi}}^{\ell}\right]
\medskip\\
& +\theta^{\mathbf{a}}\wedge\theta^{\mathbf{b}}\left[  \frac{1}{2}\left(
\nabla_{\mathbf{e}_{\mathbf{b}}}A\right)  \omega_{\mathbf{e}_{\mathbf{a}}%
}1_{\mathbf{\Xi}}^{\ell}+\frac{1}{2}\nabla_{\mathbf{e}_{\mathbf{a}}}%
^{s}\left(  A\omega_{\mathbf{e}_{\mathbf{b}}}1_{\mathbf{\Xi}}^{\ell}\right)
-\frac{1}{2}\omega_{\mathbf{ab}}^{\mathbf{c}}A\omega_{\mathbf{e}_{\mathbf{c}}%
}1_{\mathbf{\Xi}}^{\ell}\right]  .
\end{array}
\label{r12}%
\end{equation}
On the other hand,%
\begin{equation}%
\begin{array}
[c]{ll}%
\nabla_{\mathbf{e}_{\mathbf{a}}}^{s}\left(  A\omega_{\mathbf{e}_{\mathbf{b}}%
}1_{\mathbf{\Xi}}^{\ell}\right)  & =\left(  \nabla_{\mathbf{e}_{\mathbf{a}}%
}A\omega_{\mathbf{e}_{\mathbf{b}}}\right)  1_{\mathbf{\Xi}}^{\ell}%
+A\omega_{\mathbf{e}_{\mathbf{b}}}\nabla_{\mathbf{e}_{\mathbf{a}}}%
^{s}1_{\mathbf{\Xi}}^{\ell}\medskip\\
& =\left(  \nabla_{\mathbf{e}_{\mathbf{a}}}A\right)  \omega_{\mathbf{e}%
_{\mathbf{b}}}1_{\mathbf{\Xi}}^{\ell}+A\left(  \nabla_{\mathbf{e}_{\mathbf{a}%
}}\omega_{\mathbf{e}_{\mathbf{b}}}\right)  1_{\mathbf{\Xi}}^{\ell}+\frac{1}%
{2}A\omega_{\mathbf{e}_{\mathbf{b}}}\omega_{\mathbf{e}_{\mathbf{a}}%
}1_{\mathbf{\Xi}}^{\ell}\medskip\\
& =\left[  \left(  \nabla_{\mathbf{e}_{\mathbf{a}}}A\right)  \omega
_{\mathbf{e}_{\mathbf{b}}}+A\left(  \nabla_{\mathbf{e}_{\mathbf{a}}}%
\omega_{\mathbf{e}_{\mathbf{b}}}\right)  +\frac{1}{2}A\omega_{\mathbf{e}%
_{\mathbf{b}}}\omega_{\mathbf{e}_{\mathbf{a}}}\right]  1_{\mathbf{\Xi}}^{\ell
}.
\end{array}
\label{r13}%
\end{equation}
Then from Eq.(\ref{r12}) \ and Eq.(\ref{r13}) we get%
\begin{equation}
\left(  \mbox{\boldmath$\partial$}^{s}\right)  ^{2}\psi=\left[
\mbox{\boldmath$\partial$}^{2}A+\eta^{\mathbf{ab}}S_{2}A+\left(
\theta^{\mathbf{a}}\wedge\theta^{\mathbf{b}}\right)  S_{2}A\right]
1_{\mathbf{\Xi}}^{\ell} \label{r14}%
\end{equation}
where%
\begin{equation}
S_{2}A=\frac{1}{2}\left(  \nabla_{\mathbf{e}_{\mathbf{b}}}A\right)
\omega_{\mathbf{e}_{\mathbf{a}}}+\frac{1}{2}\left(  \nabla_{\mathbf{e}%
_{\mathbf{a}}}A\right)  \omega_{\mathbf{e}_{\mathbf{b}}}+\frac{1}{2}A\left(
\nabla_{\mathbf{e}_{\mathbf{a}}}\omega_{\mathbf{e}_{\mathbf{b}}}\right)
+\frac{1}{4}A\omega_{\mathbf{e}_{\mathbf{b}}}\omega_{\mathbf{e}_{\mathbf{a}}%
}-\frac{1}{2}\omega_{\mathbf{ab}}^{\mathbf{c}}A\omega_{\mathbf{e}_{\mathbf{c}%
}}. \label{r15}%
\end{equation}

Now taking into account Eq.(\ref{p4}) and the Eq.(\ref{r14}) we can write the
following relation between $\left(  \mbox{\boldmath$\partial$}^{s}\right)
^{2}$ and $%
\bpartial
^{2}$ :%
\begin{equation}
\left(  \mbox{\boldmath$\partial$}^{s}\right)  ^{2}\psi=\left[  \left(
\bpartial
\right)  ^{2}A+\eta^{\mathbf{ab}}\left(  S_{1}+S_{2}\right)  A+\left(
\theta^{\mathbf{a}}\wedge\theta^{\mathbf{b}}\right)  \left(  S_{1}%
+S_{2}\right)  A\right]  1_{\Xi}^{\ell}, \label{r16}%
\end{equation}
or
\[
\left(  \left(  \mbox{\boldmath$\partial$}^{s}\right)  ^{2}\psi\right)
1_{\Xi}^{r}=\left[  \left(
\bpartial
\right)  ^{2}A+\eta^{\mathbf{ab}}\left(  S_{1}+S_{2}\right)  A+\left(
\theta^{\mathbf{a}}\wedge\theta^{\mathbf{b}}\right)  \left(  S_{1}%
+S_{2}\right)  A\right]  .
\]
where $S_{1}$ is given by the Eq. (\ref{p5}) and $S_{2}A$ can be written in
terms of the Levi-Civita connection as%
\begin{equation}%
\begin{array}
[c]{cc}%
S_{2}A & =\frac{1}{2}\left(  D_{\mathbf{e}_{\mathbf{b}}}A\right)
\omega_{\mathbf{e}_{\mathbf{a}}}+\frac{1}{2}\left(  D_{\mathbf{e}_{\mathbf{a}%
}}A\right)  \omega_{\mathbf{e}_{\mathbf{b}}}+\frac{1}{2}A\left(
D_{\mathbf{e}_{\mathbf{a}}}\omega_{\mathbf{e}_{\mathbf{b}}}\right)  +\frac
{1}{4}A\omega_{\mathbf{e}_{\mathbf{b}}}\omega_{\mathbf{e}_{\mathbf{a}}%
}\medskip\\
& -\frac{1}{2}\omega_{\mathbf{ab}}^{\mathbf{c}}A\omega_{\mathbf{e}%
_{\mathbf{c}}}+\frac{1}{4}\tau\left(  \mathbf{e}_{\mathbf{b}},A\right)
\omega_{\mathbf{e}_{\mathbf{a}}}+\frac{1}{4}\tau\left(  \mathbf{e}%
_{\mathbf{a}},A\right)  \omega_{\mathbf{e}_{\mathbf{b}}}+\frac{1}{4}%
A\tau\left(  \mathbf{e}_{\mathbf{a}},\omega_{\mathbf{e}_{\mathbf{a}}}\right)
\omega_{\mathbf{e}_{\mathbf{b}}}.
\end{array}
\label{r17}%
\end{equation}
Notice that \ in the above formulas, the action of $\left(
\mbox{\boldmath$\partial$}^{s}\right)  ^{2}$ is on $\sec\mathcal{C}%
\ell_{\mathrm{Spin}_{1,3}^{e}}^{\ell}(M,\mathtt{g})$ and the action of the
$\mbox{\boldmath$\partial$}^{2}$ and $%
\bpartial
^{2}$ are on $\sec TM\hookrightarrow\sec\mathcal{C}\ell(M,\mathtt{g})$.

\section{Summary}

In this paper we studied the theory of the Dirac and spin-Dirac operators on
Riemann-Cartan space(time) and on a Riemannian (Lorentzian) space(time) and
introduce mathematical methods permitting the calculation of the square of
these operators, playing important role in several important topics of modern
Mathematics (in particular in the study of the geometry of moduli spaces of a
class of black holes, the geometry of NS-5 brane solutions of type II
supergravity theories and BPS solitons in some string theories) in a very
simple way. We obtain a generalized Lichnerowicz formula, and several useful
decomposition formulas for the Dirac and spin-Dirac operators in terms of the
\textit{standard }Dirac and spin-Dirac operators. Also, using the fact that
spinor fields (sections of a spin-Clifford bundle) have representatives in the
Clifford bundle we found a noticeable relation involving the spin-Dirac and
the Dirac operators, which may be eventually useful in theories using superfields.

\end{document}